\documentclass[11pt,letter]{article}

\usepackage{times}
\usepackage{graphicx}
\usepackage{subfigure}
\usepackage{enumitem}
\usepackage[hyphens]{url}
\usepackage{amsfonts,amsmath,amsthm, amssymb}
\usepackage{algorithmic, algorithm}

\usepackage[semicolon]{natbib}
\usepackage[margin=1in]{geometry}

\usepackage[usenames,dvipsnames]{xcolor}
\usepackage{balance}


\newcounter{parentnumber}

\newtheorem{theorem}{Theorem}[section]
\newtheorem{lemma}[theorem]{Lemma}
\newtheorem{proposition}[theorem]{Proposition}
\newtheorem{corollary}[theorem]{Corollary}

\newtheorem{definition}{Definition}

\newcommand{\utilityfn}{U}

\newcommand{\maxpayproofs}{\ensuremath{\alpha}}
\newcommand{\minpay}{\ensuremath{\alpha_{\min}}}
\newcommand{\maxpay}{\ensuremath{\alpha_{\max}}}

\newcommand{\numchoices}{\ensuremath{B}}
\newcommand{\numgold}{G}
\newcommand{\numques}{N}

\newcommand{\thresholdEvery}{\sigma}

\newcommand{\paysymbol}{f}

\newcommand{\pay}[1]{\ensuremath{\paysymbol(#1)}}

\newcommand{\ans}[1]{\ensuremath{x}_{#1}}
\newcommand{\anssym}{\ensuremath{x}}
\newcommand{\attemptsym}{\ensuremath{y}}
\newcommand{\attempt}[1]{\ensuremath{y}_{#1}}
\newcommand{\Epay}{\ensuremath{\mathcal{\$}}}
\newcommand{\minbelief}{\ensuremath{\rho}}
\newcommand{\skipsym}{\numchoices}
\newcommand{\const}{c}
\newcommand{\probopt}[2]{p_{#1 #2}}
\newcommand{\probques}[1]{s_{#1}}
\newcommand{\identity}[1]{\mathbf{1}\{#1\}}
\newcommand{\expect}[1]{\mathbb{E}[ #1 ]}

\newcommand{\cardinality}[1]{|#1|}
\newcommand{\defn}{:=}
\newcommand{\indicator}[1]{\mathbf{1}\{#1\}}
\newcommand{\suchthat}{\ensuremath{\mid}}
\DeclareMathOperator*{\argmax}{\mbox{arg\,max}}

\newcommand{\absolute}[1]{|#1|}

\newcommand{\selmin}{\ensuremath{s_{\min}}}
\newcommand{\selmax}{\ensuremath{s_{\max}}}

\newcommand{\arxiv}[1]{#1}
\newcommand{\icml}[1]{}



\newcommand{\citett}{\citet}
\newcommand{\citepp}{\citep}
\begin{document}
\title{Approval Voting and Incentives in Crowdsourcing}

\date{}
\maketitle
\vspace{-.6in}
\begin{center}
\large
\begin{tabular}{ccc}
Nihar B. Shah & Dengyong Zhou & Yuval Peres\\
\normalsize UC Berkeley & \normalsize Microsoft Research & \normalsize Microsoft Research\\
\normalsize \tt nihar@eecs.berkeley.edu & \normalsize \tt dengyong.zhou@microsoft.com  & \normalsize \tt peres@microsoft.com
\end{tabular}\qquad\qquad
\end{center}
\vspace{.4cm}

\begin{abstract}
The growing need for labeled training data has made crowdsourcing an important part of machine learning. The quality of crowdsourced labels is, however,
adversely affected by three factors: (1)~the workers are not experts; (2)~the incentives of the workers are not aligned with those of the requesters; and (3)~the interface does not allow workers to convey their knowledge accurately, by forcing them to make a single choice among a set of options. In this paper,  we address these issues by introducing approval voting to 
utilize the expertise of workers who have partial knowledge of the true answer, and coupling it with a (``strictly proper'') incentive-compatible compensation mechanism. We show rigorous theoretical guarantees of optimality of our mechanism together with a simple axiomatic characterization.  We also conduct preliminary empirical studies on Amazon Mechanical Turk which validate our approach.
\end{abstract}

\section{Introduction}\label{sec:introduction}

\begin{figure}
\centering
\subfigure[]{
\frame{\includegraphics[width=.29\textwidth]{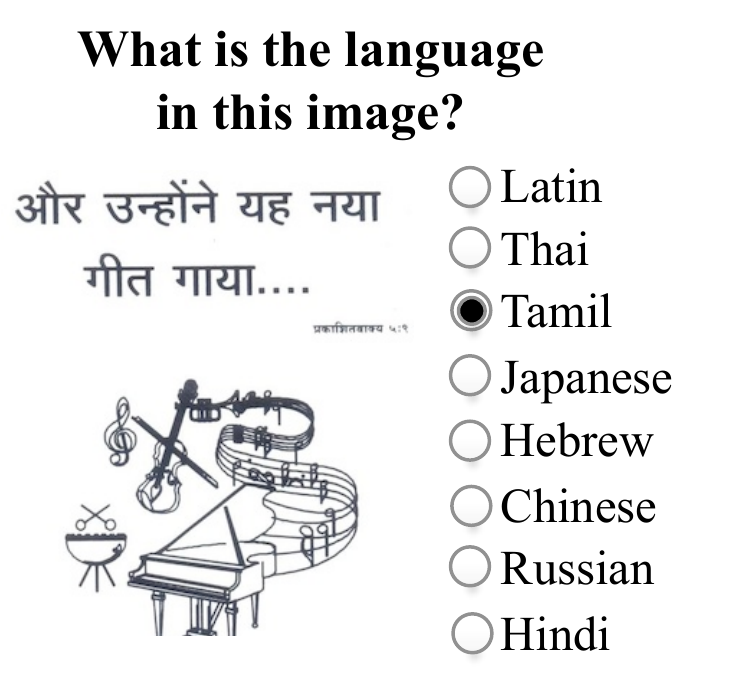}}
\label{fig:subset_intro_single}
}
\qquad \qquad
\subfigure[]{
\frame{\includegraphics[width=.32\textwidth]{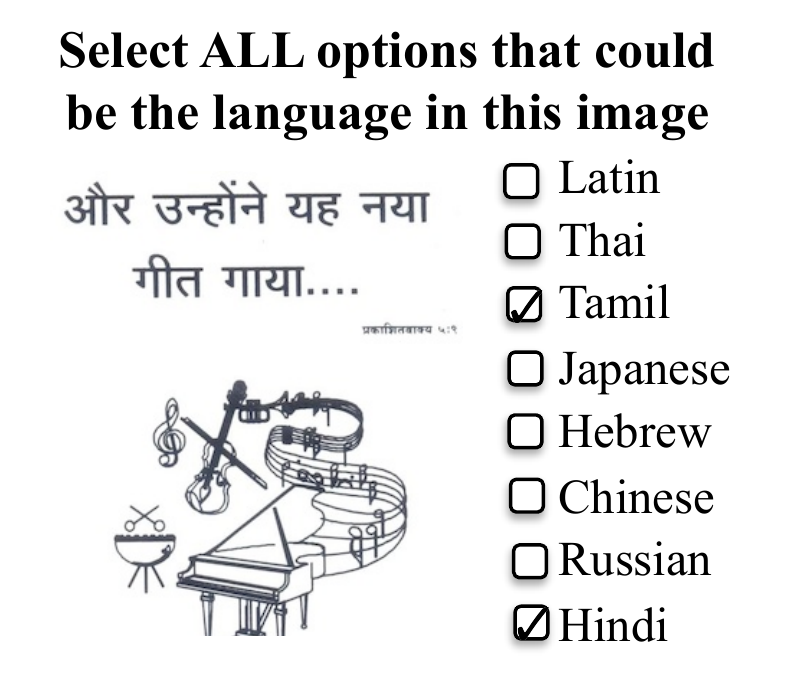}}
\label{fig:subset_intro_approval}
}
\caption{Illustration of a task with (a) the standard single selection interface, and (b) an approval-voting interface.}
\label{fig:subset_intro}
\end{figure}


In the big data era, with the ever increasing complexity of  machine learning models such as deep learning, the demand for large amounts of labeled data is growing at an unprecedented scale. A primary means of label collection is crowdsourcing, through commercial web services like Amazon Mechanical Turk where crowdsourcing workers or annotators perform tasks in exchange for monetary payments.  Unfortunately, the data obtained via crowdsourcing is typically highly erroneous~\citepp{kazai2011crowdsourcing,vuurens2011much,wais2010towards} due to the lack of expertise of workers, lack of appropriate incentives, and often the lack of an appropriate interface for the workers to express their knowledge. Several statistical aggregation methods~\citepp{dawid1979maximum,whitehill2009whose,raykar2010learning, karger2011iterative,liu2012variational,zhou2012learning, shah2015topology} have been proposed in the literature for improving the quality of the data. Our approach complements these techniques in that we endeavor to obtain higher-quality labels directly via novel interface and incentive mechanisms while not increasing the labeling cost.

The typical crowdsourcing labeling task consists of a set of questions such as images to be labeled, and each question is associated with a set of options. Each option is the name of a category and the true label for any question is one of these options. In principle, for each question, the worker is required to select the option that she believes is most likely to be correct. More formally, it involves eliciting the \emph{mode} of the worker's belief.  
Such a ``single-selection'' crowdsourcing setting has been studied extensively, both empirically and theoretically.

In this paper, we consider an alternative ``approval-voting'' means of eliciting labels from the workers, wherein the worker is allowed to select multiple options for every question.\footnote{The literature on psychology often refers to approval voting as ``subset selection''.} See Figure~\ref{fig:subset_intro} for an example. Approval voting is known to have many advantages over single-selection systems in psychology and social choice theory~\citepp{horst1932chance,coombs1953use,coombs1956assessment,collet1971elimination,brams1978approval,gibbons1979subset}: it provides workers more flexibility to express their beliefs, and utilizes the expertise of workers with partial knowledge more effectively. For instance, \citett{coombs1953use} posits that ``It seems to be a common experience of individuals taking objective tests to feel confident about eliminating some of the wrong alternatives and then guess from among the remaining ones'' and that ``Individuals taking the test should be instructed to cross out all the alternatives which they consider wrong.''  Under this approval-voting interface, we will require a worker to select every option which she believes could possibly be correct. Mathematically, we formulate this problem as eliciting the \emph{support} of the beliefs of workers for each question. In the setting of crowdsourcing, as compared to single-selection, selecting multiple options would allow for obtaining more information about the partial knowledge of these non-expert workers. This additional information is particularly valuable for difficult labeling questions, allowing for the identification of the sources of difficulty.  Indeed,~\citett{coombs1956assessment} conclude that under such a questionnaire, ``clear evidence for the existence of partial information mediating responses to multiple choice items was obtained.''

\arxiv{
Let us illustrate the utility of approval voting using an example in Figure~\ref{fig:subset_intro}. Assume that there are two workers. The first worker believes the true label to be either ``cheetah''  or ``leopard'', but certainly not any other option; the second worker is confused about some other aspect of the image, and believes the true label to be either  ``cheetah'' or ``jaguar'', but certainly none of the others. If each worker is allowed to select only a single answer, it may turn out that the first worker selects ``leopard'' and the second worker selects ``jaguar''. Their responses will thus not provide any definitive answer about the true label. In contrast, if we fully elicit their knowledge by letting them select multiple options, that is, (``cheetah'',  ``leopard'') from the first worker and (``cheetah'', ``jaguar'') from the other worker, then ``cheetah'' becomes a clear winner.} 


Albeit its great flexibility in eliciting partial knowledge, approval  voting alone is not sufficient for high quality crowdsourcing. A worker may have no incentive to truthfully disclose her partial knowledge on the crowdsourcing question. For instance, the worker may simply choose all provided options as her answer and get paid. To address this problem, we need to couple approval voting with an appropriate ``incentive-compatible'' payment mechanism such that a worker receives her maximum expected payment if and only if she truthfully discloses her partial knowledge (that is, the support of her belief) on the crowdsourcing question. In other words, the payment mechanism has to be a ``strictly proper scoring rule''. Moreover, we want the mechanism to be ``frugal'', paying as less as possible to a worker who simply selects all provided options as her answer. The problem setting for incentive mechanism design is formally described in Section~\ref{sec:problem}.

Our first result is negative, proving that unfortunately no mechanism can be incentive compatible for this setting (Section~\ref{sec:impossible}). This impossibility result leads us to introduce a ``coarse belief'' assumption that relies on a certain granularity in people's beliefs. 

Our next result is the design of a payment mechanism and associated proofs showing that our mechanism is incentive compatible and frugal (Section~\ref{sec:mechanism}). Furthermore, we show that it is the only mechanism which satisfies these two requirements. 

We then generalize the analysis of our mechanism to settings where the coarse belief assumption may not be satisfied, and show that our mechanism simply incentivizes workers to select options for which their belief is relatively high enough (Section~\ref{sec:properties}). This perspective also leads to a simple axiomatic characterization of our mechanism. 

We then report results from preliminary experiments verifying certain basic hypotheses underlying our approach (Section~\ref{sec:experiments}). The paper then diversifies to investigate two closely related settings, that of general utility functions (Section~\ref{sec:utility}) and that of a problem of reporting only high enough beliefs (Section~\ref{sec:thresholdEvery}). The paper concludes with a discussion in Section~\ref{sec:conclusion}.

\subsection*{Related literature}
Approval voting~\citepp{ottewell1977arithmetic,kellett1977presidential,weber1977comparison,brams1978approval} is a form of voting in which each voter can ``approve of'' (that is, select) multiple candidates. No further preferences among these candidates is specified by the voter. Our proposed interface for crowdsourcing elicits approvals on the candidate options for each question. Closer to our setting of crowdsourcing, approval voting has been studied in the context of question and answer forums~\citepp{jain2009designing} and Doodle polls~\citepp{zou2014approval}. The focus of the present paper is on the design of incentive mechanisms with properties that fundamentally hold irrespective of the nature of the setting.

The framework of scoring rules~\citepp{brier1950verification,savage1971elicitation,gneiting2007strictly,lambert2009eliciting}  considers the design of payment mechanisms to elicit predictions about an event whose actual outcome will be observed in the future. The payment is a function of the agent's response and the outcome of the event. The payment is called ``strictly proper'' if its expectation, with respect to the belief of the agent about the event, is strictly maximized when the agent reports her true belief. Proper scoring rules however provide a very broad class of mechanisms, and \emph{do not specify} any particular mechanism for use. The mechanism proposed in the present paper may alternatively be viewed as the ``optimal'' proper scoring rules for eliciting supports of workers' beliefs across multiple questions.

\citett{shah2015double} consider a crowdsourcing setup with the traditional single-selection setting, also eliciting the workers' confidences for each response. They propose a mechanism to suitably incentivize workers and show that their proposed mechanism is shown to be the only one satisfying a proposed ``no-free-lunch'' axiom. While the setting of our work is different from that of~\citett{shah2015double}, interestingly, our mechanism that was derived for a different interface and under a different set of assumptions, turns out to be the only mechanism that can satisfy the no-free-lunch axiom (adapted to our setting).

The mechanisms presented subsequently in the present paper assume the presence of some ``gold standard'' questions whose answers are known apriori to the system designer. There is a parallel line of literature~\citepp{prelec2004bayesian,miller2005eliciting,faltings2014incentives,miller2005eliciting,dasgupta2013crowdsourced} that explores the design of mechanisms that operate in the absence of any gold standard questions. These works typically elicit additional information from the workers, such as asking them to predict the responses of other workers. The mechanisms designed therein can generally provide only weaker guarantees due to the absence of a gold standard answer to compare with.


\section{Problem setup}\label{sec:problem}

Consider $\numques \geq 1$ questions, each of which has $\numchoices \geq 2$ options to choose from. For each option, exactly one of the $\numchoices$ options is correct. We assume that these $\numques$ questions contain $\numgold~(1 \leq \numgold \leq \numques)$ ``gold standard'' questions, that is, questions to which the mechanism designer knows the answers apriori. These gold standard questions are assumed to be mixed uniformly at random among the $\numques$ questions, and the worker is evaluated based on her performance on these $\numgold$ questions. For every individual question, we assume that the worker has, in her mind, a distribution over the $\numchoices$ options representing her beliefs of the probabilities of the respective options being correct. We assume that these belief-distributions of a worker are independent across questions~\citepp{gibbons1979subset}. For any integer $K$, we will use the standard notation of $[K]$ as a shorthand for the set $\{1,\ldots,K\}$.

Our goal is to elicit, for every question, the \emph{support} of the worker's distribution over the $\numchoices$ options. In other words, we wish to incentivize the worker such that for each question, the worker should select the \emph{smallest subset of the set of options} such that the correct answer according to her belief lies in the selected subset. Formally, suppose that for any question $i \in [\numques]$, the worker believes that the probability of option $b \in [\numchoices]$ being correct is $\probopt{i}{b}$, for some non-negative values $\probopt{i}{1},\ldots,\probopt{i}{\numchoices}$ that sum to one. Then the goal is to incentivize the worker to, for each question $i \in [\numques]$, select precisely the set of options
\begin{align}
\{b \in [\numchoices] \mid \probopt{i}{b} \neq 0\}.
\label{eq:support_definition}
\end{align}

\textbf{Payment function.} As mentioned earlier, the worker's performance is evaluated based on her responses to the gold standard questions. For any question in the gold standard, we denote the evaluation of the worker's performance on this question by a value in the set $\{-(\numchoices-1),\ldots,-1,1,\ldots,\numchoices\}$: the magnitude of this value represents the number of options she had selected and the sign is positive if the correct answer was in that subset and negative otherwise. For instance, if the worker selected four options for a certain gold standard question but none of them was correct, then the evaluation of this response is denoted as ``$-4$''; if the worker selects two options for a gold standard question and one of them turns out to be the correct option then the evaluation of this response is denoted as ``$+2$''. 

We will assume that the payments are bounded, that is, any payment must lie in the interval $[\minpay, \maxpay]$, for some values $\minpay$ and $\maxpay > \minpay$. The choice of the two parameters $\minpay$ and $\maxpay$ may be made keeping various factors in mind, such as guidelines of the crowdsourcing platform used, the budget constraints, and the minimum wage. We will assume that the values of the two parameters are given to us.

Let 
\begin{align*}\paysymbol:\{-(\numchoices-1),\ldots,-1,1,\ldots,\numchoices\}^\numgold \rightarrow [\minpay, \maxpay]
\end{align*} 
denote the payment function. It is this function $\paysymbol$ which must be designed in order to incentivize the worker. 

We will let that a worker who answers everything perfectly should be paid an amount $\maxpay$, that is,
\begin{align}
\pay{1,\ldots,1} = \maxpay.~\label{eq:maxpay}
\end{align}

\textbf{Expected payment.} A quantity central to our analysis is the \emph{expected payment}, where the expectation is from the point of view of the worker, and is taken over the randomness in the choice of the $\numgold$ gold standard questions among the $\numques$ questions, and over the $\numques$ probability distributions representing her beliefs for the $\numques$ questions. Let us formalize this notion. Suppose that for question $i \in [\numques]$, the worker has selected some $\attempt{i} \in [\numchoices]$  of the $\numchoices$ options. Further, let $\probques{i} \in [0,1]$ denote the probability, under the worker's beliefs, that the correct answer to question $i$ lies in this set of $\attempt{i}$ selected options. In other words, $\probques{i}$ denotes the sum of the beliefs for the $\attempt{i}$ options selected by the worker (consequently, the sum of the beliefs for the options not selected is $(1-\probques{i})$). Then from the worker's point of view, her expected payment for this selection is
\begin{align}
\frac{1}{{\numques \choose \numgold}}  \sum_{\substack{(j_1,\ldots,j_\numgold) \subseteq [\numques]}} \sum_{\substack{(\epsilon_1,\ldots,\epsilon_\numgold )  \in \{-1,1\}^\numgold }}  \Big(   \prod_{i=1}^{\numgold} & (1-\probques{j_i})^{\identity{\epsilon_i=-1}} \probques{j_i}^{\identity{\epsilon_i=1}}
 \pay{\epsilon_1 \attempt{j_1},\ldots, \epsilon_\numgold \attempt{j_\numgold}} \Big) .
\label{eq:epay_definition}
\end{align}

The outer summation in~\eqref{eq:epay_definition} corresponds to the expectation with respect to the random distribution of the $\numgold$ gold standard questions in the $\numques$ total questions, and the inner summation corresponds to the expectation with respect to the worker's beliefs of her choices being correct. In this paper, we assume that the workers aim to maximize their expected rewards; extending our theory to more general utility functions is straightforward.

Given the presence of gold standard questions, the performance of any worker is based only on her responses to questions to which answers are already known by the mechanism designer, the payments made to different workers do not depend on each other and hence we consider only one worker without loss of generality.

\textbf{Goal.} The goal is to design mechanisms that are incentive compatible:
\begin{definition}[Incentive compatibility]
A mechanism  is incentive compatible if the expected payment (Equation~\eqref{eq:epay_definition}), from the worker's point of view, is strictly maximized when she selects precisely the support (Equation~\eqref{eq:support_definition}) of her belief for each question.
\end{definition}
Note that the definition of incentive compatibility used here considers a ``strict'' maximization.

Observe that a worker who selects all the options for all the questions doesn't give any useful information. In order to deter such ``freeloading'' behavior, one would like to ensure that in addition to paying a (large enough) amount $\maxpayproofs$ to a good worker, the mechanism should expend as small an amount as possible on such a worker. This leads to a notion of ``frugality''.
\begin{definition}[Frugality]
An incentive-compatible mechanism $\paysymbol$ is frugal if
\begin{align*}
\pay{\numchoices,\ldots,\numchoices} \leq \paysymbol'(\numchoices,\ldots,\numchoices)	
\end{align*}
for every incentive-compatible mechanism $\paysymbol'$ that has $\paysymbol'(1,\ldots,1)=\paysymbol(1,\ldots,1)$.
\end{definition}

Our goal is to design mechanisms that are incentive-compatible, and whenever they exist, find the mechanism(s) that is (are) most frugal.



\section{An impossibility result and a coarse-beliefs assumption}\label{sec:impossible}
It turns out that, unfortunately, we must face a roadblock in the first step: We can show that there exists no mechanism that is incentive compatible.
\begin{theorem}\label{thm:support_impossible}
For any $\numques$, $\numgold$ and $\numchoices \geq 2$, there is no mechanism that can guarantee that the worker will be incentivized to select precisely the support of her distribution for each question.
\end{theorem}
The proof of this result and other theoretical results (except Theorem~\ref{thm:working}) are provided in the appendix.

In order to circumvent this impossibility result, we appeal to a certain well-understood property of human belief.

\subsection*{Coarse beliefs assumption}
There is an extensive literature in psychology establishing the coarseness of processing and perception in humans. For instance, Miller's celebrated paper~\citepp{miller1956magical} establishes the information and storage capacity of humans, that an average human being can typically distinguish at most about seven states. This granualrity of human computation is verified in many subsequent experiments~\citepp{shiffrin1994seven,saaty2003magic}. \citett{jones2013optimal} establish the ineffectiveness of finer-granularity response elicitation. \citett{mullainathan2008coarse} hypothesize that humans often group things into categories; this hypothesis is experimentally verified by~\citett{siddiqi2011does} in a specific setting. We incorporate this established notion of coarseness of human processing in our model in terms of a simple assumption.

Consider some (fixed and known) value $\minbelief>0$, and assume that the probability of any option for any question, according to the worker's belief, is either zero or greater than $\minbelief$. The impossibility shown in Theorem~\ref{thm:support_impossible} pertains to $\minbelief=0$. Also, one must necessarily take into account situations when a worker is totally clueless about a question, that is, when her belief is distributed uniformly over all options. Hence we restrict $\minbelief < \frac{1}{\numchoices}$. To summarize, we make the following ``coarse belief'' assumption.
\begin{definition}[Coarse belief assumption]
The worker's belief for any option for any question lies in the set $\{0\} \cup (\minbelief,1]$ for some (fixed and known) $\minbelief \in \left(0, \frac{1}{\numchoices}\right)$.
\end{definition}


We wish to elicit the full support of the workers' beliefs, given a coarseness of belief that assigns a value of zero to very low probability categories. The goal is to design mechanisms that are incentive-compatible and frugal, assuming the coarse belief assumption holds true.

\section{Incentive mechanism}\label{sec:mechanism}
Mechanism~\ref{algo:incentive_minbelief} presents our proposed mechanism for the problem at hand, under the coarse belief assumption.
\begin{algorithm}[h!]
 \floatname{algorithm}{Mechanism}
\begin{itemize}[leftmargin=*]
\item \textbf{Input:} Evaluations of the worker's answers to the $\numgold$ gold standard questions\\ $(\ans{1},\ldots,\ans{\numgold}) \in \{-(\numchoices-1),\ldots,-1,1,\ldots,\numchoices\}^\numgold$
\item \textbf{Output:} The worker's payment
\begin{equation*}
f(\ans{1},\ldots,\ans{\numgold}) = (\maxpay-\minpay) (1-\minbelief)^{\sum_{i = 1}^G (\ans{i}-1)} \prod_{i = 1}^G\mathbf{1}\{\ans{i} \geq 1\} + \minpay
\end{equation*}
\end{itemize}
\caption{Incentive mechanism for approval voting}
\label{algo:incentive_minbelief}
\end{algorithm}

The payment is based only on the evaluation of the worker's responses to the gold standard questions. It is easy to describe the mechanism in words: The payment is $\minpay$ plus
 \begin{itemize}
 \item $0$ if the correct answer is not selected for any of the questions, otherwise
 \item $(\maxpay - \minpay)$ reduced by $ (100\minbelief) \%$ for each incorrect option selected.
 \end{itemize}

The following pair of theorems present our main results, proving that this mechanism is \emph{the one and only} mechanism that satisfies our requirements.
\begin{theorem}\label{thm:working}
Under the coarse-beliefs assumption, Mechanism~\ref{algo:incentive_minbelief} is incentive-compatible and frugal.
\end{theorem}

\noindent The following theorem shows that our mechanism is \textit{strictly} better than any other mechanism.
\begin{theorem}\label{thm:unique}
Under the coarse-beliefs assumption, there is no other incentive-compatible mechanism that expends as small an amount as Mechanism~\ref{algo:incentive_minbelief} on a worker who does not attempt any question.
\end{theorem}

To show the optimality and uniqueness properties claimed in Theorem~\ref{thm:working} and Theorem~\ref{thm:unique} respectively, we prove the absence of other good mechanisms via contradiction-based arguments. Specifically, for any candidate mechanism, we identify a set of beliefs for which the worker will not be incentivized to act as required. In line with our earlier argument of beliefs being ``coarse'', the beliefs considered in these proofs are simple enough: the worker has some belief about one of the options, knows for sure that certain other options are incorrect, and is indifferent among the rest of the options.

To put things in perspective, observe that $\minbelief=0$ eliminates the dependence of the payment in Mechanism~\ref{algo:incentive_minbelief} on $\sum_i \ans{i}$ and makes the mechanism incentive incompatible. The impossibility result of Theorem~\ref{thm:support_impossible} proves that every possible mechanism must necessarily suffer this fate.


The remainder of this section is devoted to the proof of Theorem~\ref{thm:working}. The reader may feel free to jump to Section~\ref{sec:properties} without any loss in continuity. 

\subsection*{Proof of Theorem~\ref{thm:working}}
Without loss of generality, assume that $\minpay = 0$ since in our setting, the property of incentive compatibility is invariant to any constant shift and positive scale of the payment. We adopt the succinct notation of $\maxpayproofs \defn \maxpay - \minpay$.

\textbf{Incentive compatibility.}
First consider the case $\numques = \numgold = 1$. In this case, Mechanism~\ref{algo:incentive_minbelief} reduces to
\begin{align*}
\pay{\anssym} = \maxpayproofs (1-\minbelief)^{(\ans{1}-1)} \mathbf{1}\{\ans{1} \geq 1\}.
\end{align*}
Suppose without loss of generality that the worker's beliefs for the $\numchoices$ options are $p_1 \geq \cdots \geq p_m > \minbelief > p_{m+1} = \cdots = p_{\numchoices} = 0$ for some $m \in [\numchoices]$. An incentive-compatible mechanism must strictly maximize the worker's expected payment when she selects the support of her belief, that is, the options $\{1,\ldots,m\}$. The expected payment, $\Epay_{\mbox{sup}}$, under this selection is
\begin{align*}
\Epay_{\mbox{sup}} & = \maxpayproofs \sum_{i=1}^{m} p_i (1-\minbelief)^{m-1} \\
& = (1-\minbelief)^{m-1}.
\end{align*}
Suppose the worker selects some other set of options $\{o_1,\ldots,o_\ell\} \subseteq [\numchoices]$, $\{o_1,\ldots,o_{\ell}\} \neq [m]$. Then her expected payment $\Epay_{\mbox{oth}}$ under the proposed mechanism for this selection is
\begin{align}
\Epay_{\mbox{oth}} &= \maxpayproofs \sum_{i=1}^{\ell} p_{o_i} (1-\minbelief)^{\ell-1} \nonumber \\
& \leq \maxpayproofs \sum_{i=1}^{\ell} p_{i} (1-\minbelief)^{\ell-1},\label{eq:working1}
\end{align}
since $p_1 \geq \cdots  \geq p_{\numchoices}$. If $\ell = m$ then the inequality in~\eqref{eq:working1} is strict since $p_j < p_i$ for all $(j>m,\,i \leq m)$. Thus the expected payment under the choice $\ell = m$ but with a selection different from the support is strictly lower than $\Epay_{\mbox{sup}}$. Also observe that the expected payment on selecting $\ell > m$ is upper bounded by $(1-\minbelief)^{\ell-1}$, which is strictly smaller than $\Epay_{\mbox{sup}}$. Let us now consider the remaining, interesting case of $\ell < m$. Since $p_i > \minbelief$ for all $i \in [m]$, we have
\begin{align*}
\Epay_{\mbox{oth}} & < \maxpayproofs \left(\sum_{i=1}^{m} p_{i} - (m-\ell)\minbelief \right) (1-\minbelief)^{\ell-1}\\
 &=  \maxpayproofs \left(1 - (m-\ell)\minbelief \right) (1-\minbelief)^{\ell-1}\\
& \leq  \maxpayproofs \left(1 - (m-(\ell+1))\minbelief \right) (1-\minbelief)^{\ell} \\
& ~\vdots \\
& \leq \maxpayproofs (1-\minbelief)^{m-1} \\
& = \Epay_{\mbox{sup}}.
\end{align*}
This completes the proof for the case $\numques = \numgold = 1$.

Let us now consider the case of $\numques = \numgold \geq 1$. By our assumption of the independence of the beliefs of the worker across the questions, the expected payment equals
\begin{align*}
\prod_{i=1}^{\numgold} \mathbf{E} \left[ \maxpayproofs (1-\minbelief)^{(\ans{i}-1)} \mathbf{1}\{\ans{i} \geq 1\} \right].
\end{align*}
Since the payments are non-negative, if each individual component in the product is maximized then the product is also necessarily maximized. Each individual component simply corresponds to the setting of $\numques = \numgold = 1$ discussed earlier. Thus calling upon our earlier result, we get that the expected payment for the case $\numques = \numgold > 1$ is maximized when the worker acts as desired for every question.

Let us finally consider the case of $\numques > \numgold \geq 1$. Recall from~\eqref{eq:epay_definition} that the expected payment for the general case is a cascade of two expectations: the outer expectation is with respect to the uniformly random distribution of the $\numgold$ gold standard questions among the $\numques$ total questions, while the inner expectation is taken over the worker's beliefs of the different questions conditioned on the choice of the gold standard questions. The arguments above for the case $\numques = \numgold$ prove that every individual term in the inner expectation is maximized when the worker acts as desired. The expected payment is thus maximized when the worker acts as desired.

\textbf{Frugality.}
We first present a lemma that forms the workhorse of this and other subsequent proofs. \begin{lemma}
Consider some $\attemptsym, \attemptsym' \in [\numchoices]^\numques$ and some $\mathcal{I} \subseteq [\numques]$ such that $\attempt{i} = \attempt{i}'+1$ for all $i \in \mathcal{I}$, and $\attempt{i} = \attempt{i}'$ for all $i \notin \mathcal{I}$. Then any incentive compatible mechanism $\paysymbol$ must necessarily satisfy
\icml{
\begin{align*}
\frac{1}{{\numques \choose \numgold}} &\!\! \sum_{(j_1,\ldots,j_\numgold) \subseteq [\numques]} \pay{\attempt{j_1},\ldots,\attempt{j_\numgold}} \\
& \!\!\!\!\!\! \geq  \frac{1}{{\numques \choose \numgold}}\!\! \sum_{(j_1,\ldots,j_\numgold) \subseteq [\numques]} \!\! \!\! \!\!\!\! (1-\minbelief)^{\cardinality{\mathcal{I} \cap \{j_1,\ldots, j_\numgold\} } } \pay{\attempt{j_1}',\ldots,\attempt{j_\numgold}'}.
\end{align*}
}
\arxiv{
\begin{align*}
\frac{1}{{\numques \choose \numgold}}  \sum_{(j_1,\ldots,j_\numgold) \subseteq [\numques]} \pay{\attempt{j_1},\ldots,\attempt{j_\numgold}}
 \geq  \frac{1}{{\numques \choose \numgold}} \sum_{(j_1,\ldots,j_\numgold) \subseteq [\numques]} (1-\minbelief)^{\cardinality{\mathcal{I} \cap \{j_1,\ldots, j_\numgold\} } } \pay{\attempt{j_1}',\ldots,\attempt{j_\numgold}'}.
\end{align*}
}
Furthermore, a necessary condition for the above equation to be satisfied with equality is
\begin{align*}
& \pay{\epsilon_1 \attempt{j_1}', \ldots, \epsilon_\numgold \attempt{j_\numgold}' } = 0
\end{align*}
for all $(j_1,\ldots,j_\numgold) \subseteq [\numques]$, and all $\{(\epsilon_1,\ldots,\epsilon_\numgold) \in \{-1,1\}^{\numgold} \backslash \{1\}^\numgold  \suchthat \epsilon_i = 1 ~\mbox{whenever}~ j_i \notin \mathcal{I} \}$.
\label{lem:continuity_minbelief}
\end{lemma}

The proof of the lemma is provided in the appendix. We now prove the frugality of our proposed mechanism using this lemma. Consider any incentive compatible mechanism $\paysymbol$ such that $\pay{1,\ldots,1} = \maxpayproofs$. Consider any $\ans{0} \in [\numchoices-1]$. Applying Lemma~\ref{lem:continuity_minbelief} with $\attemptsym = (\ans{0}+1,\ldots,\ans{0}+1)$, $\attemptsym' = (\ans{0},\ldots,\ans{0})$ and $\mathcal{I} = [\numgold]$ gives
\begin{align*}
\pay{\ans{0}+1,\ldots,\ans{0}+1} \geq (1-\minbelief)^\numgold \pay{\ans{0},\ldots,\ans{0}}.
\end{align*}
A repeated application of this inequality for all $\ans{0} \in [\numchoices-1]$ gives
\begin{align*}
\pay{\numchoices,\ldots,\numchoices} & \geq (1-\minbelief)^{(\numchoices-1)\numgold} \pay{1,\ldots,1}\\
& = (1-\minbelief)^{(\numchoices-1)\numgold} \maxpayproofs.
\end{align*}
Mechanism~\ref{algo:incentive_minbelief} achieves this lower bound on $\pay{\numchoices,\ldots,\numchoices}$ with equality, thereby completing the proof.

\section{Robustness to the coarse beliefs assumption}\label{sec:properties}
We earlier made the ``coarse belief'' assumption that the worker's belief for any option, when non-zero, is atleast $\minbelief$. We then designed the Mechanism~\ref{algo:incentive_minbelief} that is incentive compatible with respect to eliciting the supports of the beliefs of the worker. A natural question then arises is: How does the mechanism perform if the coarse beliefs assumption is violated? Does the mechanism break down?

In this section, we generalize the results presented earlier in the paper to the setting where workers may have arbitrary beliefs. It turns out that our proposed mechanism continues to incentivize workers to act in a certain desirable way.

\subsection{Incentivizing workers with finer beliefs}
Suppose that Mechanism~\ref{algo:incentive_minbelief} (for a certain value of $\minbelief$) is encountered by a worker who may have arbitrary beliefs. Interestingly, it turns out that the mechanism doesn't break down, but instead does something desirable: it incentivizes the worker to select all options for which the relative belief of the worker is high enough.
\begin{theorem}
Under Mechanism~\ref{algo:incentive_minbelief}, for any question, a worker with beliefs $1 \geq p_1 \geq \ldots \geq p_\numchoices \geq 0$ will be incentivized to select options $\{1,\ldots,m\}$ where 
\begin{align*}
m = \argmax_{z \in [m]} \left( \frac{p_{z}}{\sum_{i=1}^{z}p_{i}} > \minbelief \right).
\end{align*}
\label{thm:IC_arbitrary}
\end{theorem}
It is not hard to interpret this incentivized action. The worker selects options one by one in decreasing order of her beliefs as long as the selected option contributes a fraction more than $\minbelief$ to the total belief of the selected options.

Let us now verify that the earlier result of Theorem~\ref{thm:working} for ``coarse beliefs'' is indeed a special case of Theorem~\ref{thm:IC_arbitrary}. To this end, suppose the beliefs of the worker for any particular question are $p_1 \geq \cdots p_k > \minbelief > p_{k+1} = \cdots = p_\numchoices = 0$ for some $k \in [\numchoices]$. Then we have
\begin{align*}
\frac{p_{z}}{\sum_{i=1}^{z}p_{i}} = \frac{0}{\sum_{i=1}^{z}p_{i}} = 0 < \minbelief \qquad \mbox{for all }z \geq k+1,
\end{align*}
and
\begin{align*}
\frac{p_{z}}{\sum_{i=1}^{z}p_{i}} \geq \frac{p_{z}}{1} > \minbelief \qquad \mbox{for all}~z \leq k.
\end{align*}
It follows that under the result of Theorem~\ref{thm:IC_arbitrary}, a worker with ``coarse beliefs'' will be incentivized to select precisely the support of her beliefs.

\subsection{An axiomatic derivation}
We now present an alternative axiomatic derivation of our mechanism when accommodating workers with arbitrary beliefs. 
%
%
The derivation involves a ``no-free-lunch axiom'' of~\citett{shah2015double}, which when adapted to our approval-voting based setting is defined as follows. We say that a worker has `attempted' a question if for that question, she doesn't select all the $\numchoices$ options. We say that the answer to a question is wrong if the correct option does not lie in the set of selected options.
\begin{definition}[No-free-lunch; adapted from~\citett{shah2015double}]
If the answer to every attempted question in the gold standard turns out to be wrong, then the worker gets a payment of zero, namely, 
\icml{
\begin{align*}
&\pay{\ans{1},\ldots,\ans{\numgold}} = 0\\
& ~~~~~~\forall~~ (\ans{1},\ldots,x_\numgold)\in \{-(\numchoices-1),\ldots,-1,\numchoices\}^\numgold \backslash \{\numchoices\}^\numgold.
\end{align*}
}
\end{definition}

\arxiv{
\begin{align*}
\pay{\ans{1},\ldots,\ans{\numgold}} = 0 ~~~~~~\forall~~ (\ans{1},\ldots,x_\numgold)\in \{-(\numchoices-1),\ldots,-1,\numchoices\}^\numgold \backslash \{\numchoices\}^\numgold.
\end{align*}
}

The no-free-lunch axiom is quantitatively different from the criterion of frugality proposed in this paper. However, both these notions have the same qualitative goal, namely to minimize the expenditure when no useful data is obtained, while providing higher payments to workers providing better data. Interestingly, as we show below, both these notions lead to the same (unique) mechanism under our setting of approval voting.

\begin{theorem}
Consider no assumptions on the minimum value of the belief, and suppose the workers must be incentivized to select options $\{1,\ldots,m\}$ where $m = \argmax_z \left( \frac{p_{z}}{\sum_{i=1}^{z}p_{i}} > \minbelief \right)$. Then, Mechanism~\ref{algo:incentive_minbelief} is the one and only mechanism that is incentive compatible and satisfies no-free-lunch.
\label{thm:nfl}
\end{theorem}


\section{Preliminary experiments}\label{sec:experiments}

\begin{figure}
\centering
\subfigure[]{
\frame{\includegraphics[width = .356 \textwidth]{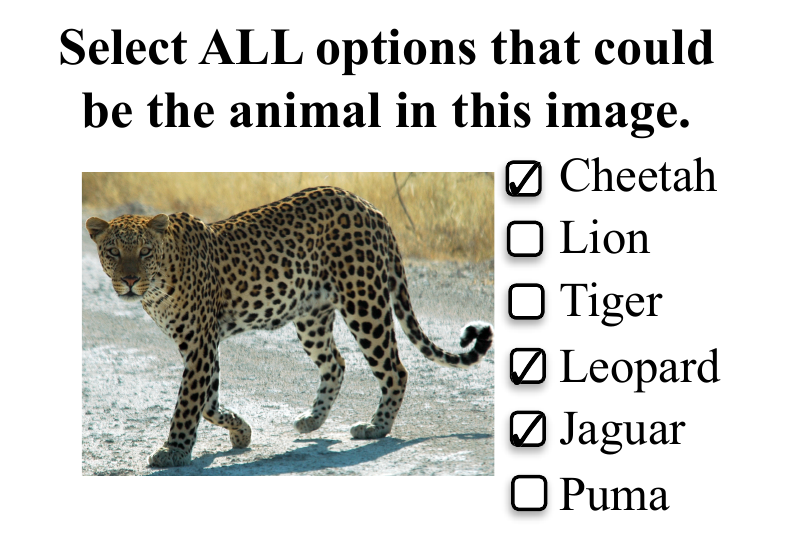}}
\label{fig:experiment_animals_approval}
}\qquad\qquad
\subfigure[]{
\frame{\includegraphics[width = .3 \textwidth]{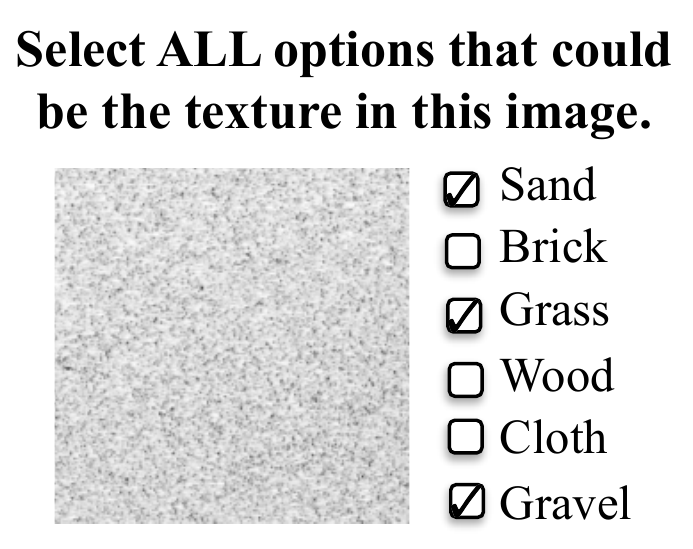}}
\label{fig:experiment_textures_approval}
}
\caption{Illustration of two of the three experiments we conducted on Amazon Mechanical Turk.}
\end{figure}

This section presents results from an evaluation of our proposed mechanism, Mechanism~\ref{algo:incentive_minbelief}, on the popular Amazon Mechanical Turk~(\url{mturk.com}) commercial crowdsourcing platform. 
The goal of this preliminary experimental exercise is to perform a basic check on whether our mechanism has the potential to work in practice. Specifically, our goal is to evaluate the primary hypotheses underlying the theory: (i) whether workers are able to make a judicious use of the approval voting setup, (ii) whether the existence of the mechanism make any difference, and (iii) if there is a opposition from the workers to the interface or the mechanism for any reason.

It is important to keep in mind that conclusive experiments for mechanism design are in general quite expensive with respect to time (workers may need months to understand a new mechanism) and budget. They are unlike typical machine-learning experiments that require only existing benchmark datasets. Moreover, the wordings or the interface may exert a significant influence on the workers' behavior. Like most mechanism design papers, we position our work primarily as a theoretical study. We expect that more detailed experiments will follow the publication of our work; indeed, it is best if experiments on such incentive schemes are conducted by multiple groups.

\subsection{Methods}

We conducted three separate sets of experiments, with over 200 workers in each experiment:
\begin{itemize}
\item Identifying languages from displayed text (Figure~\ref{fig:subset_intro})
\item Identifying animals in displayed images (Figure~\ref{fig:experiment_animals_approval})
\item Identifying textures in displayed images (Figure~\ref{fig:experiment_textures_approval}).
\end{itemize}
In each experiment, every worker was assigned one of four mechanisms uniformly at random. The variable component of each mechanism was executed as a ``bonus payment'' based on the evaluation of the worker's performance on the gold standard questions, on top of a guaranteed payment of $10$ cents (this was $\minpay$). The four mechanisms tested were:
\begin{itemize}
\item Single-selection interface with additive payments: The worker must select a single option for every question. The bonus starts at zero and is increased additively by a fixed amount for every correct answer.
\item Skip-based single-selection interface with multiplicative payments~\citepp{shah2015double}: For every question, the worker can either select one option or skip the question. The bonus starts at a certain positive value, is reduced by a certain fraction for each skipped question, and becomes zero in case of an incorrect answer.
\item Approval-voting interface with a fixed payment: The bonus is fixed.
\item Approval-voting interface with Mechanism~\ref{algo:incentive_minbelief}.
\end{itemize}
Given the caveats associated to experiments on mechanism design as mentioned earlier, we provided detailed instructions about the task and the mechanism to each worker, and also made them work through multiple examples. The entire data related to the experiments, including the interfaces used, specifics about the payment mechanisms, and the responses of the workers, is available on the website of the first author.

\subsection{Results}
\begin{figure*}
\centering
\subfigure[Fraction of responses that evaluate to different values. The magnitude of the evaluation represents the number of options selected and its sign denotes whether the correct option was selected (positive) or not (negative).]{
\includegraphics[width = .98\textwidth]{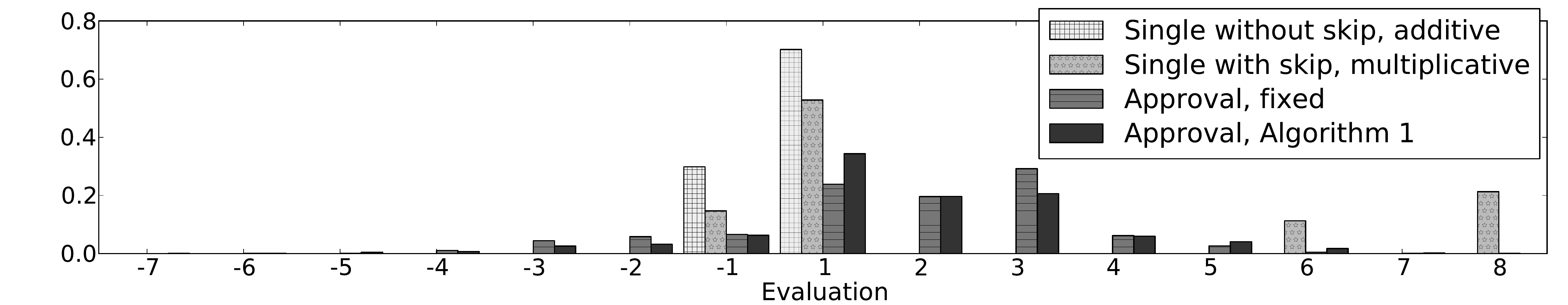}
\label{fig:plots_breakup}
}~\\
\subfigure[Fraction wrong among attempted questions]{
\qquad
\includegraphics[width = .15\textwidth]{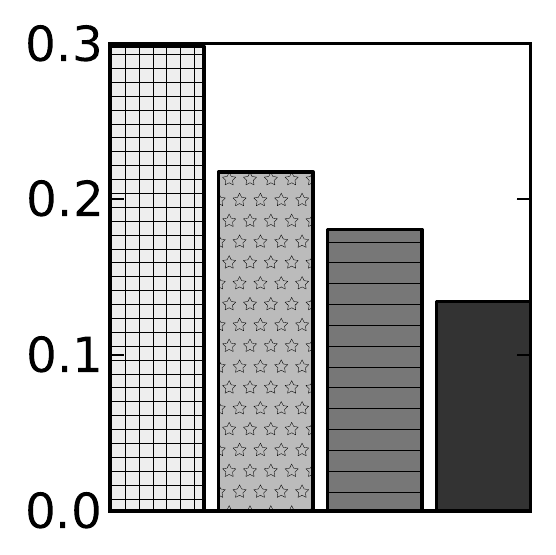}
\qquad
\label{fig:plots_nonskip}
}\qquad\qquad
\subfigure[Fraction wrong when only one option was selected]{
\qquad
\includegraphics[width = .15\textwidth]{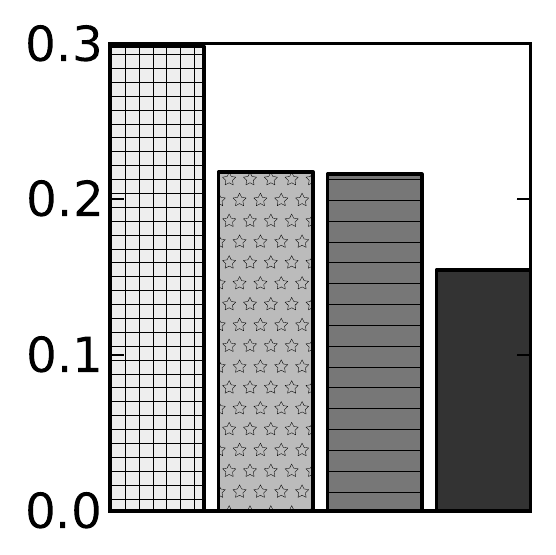}
\qquad
\label{fig:plots_one}
}\qquad\qquad
\subfigure[Average bonus per worker (cents)]{
\qquad
\includegraphics[width = .15\textwidth]{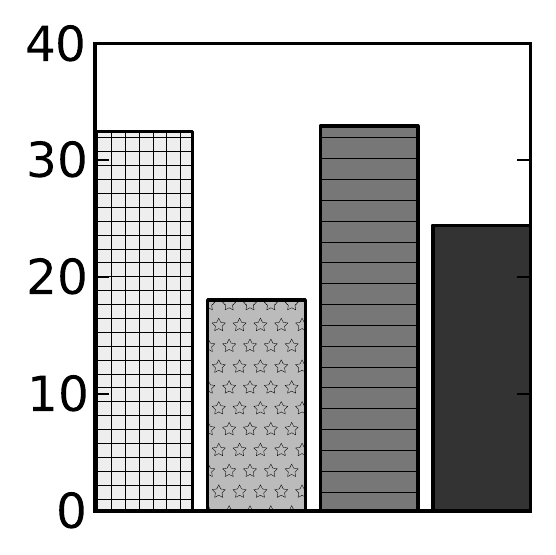}
\qquad
\label{fig:plots_payment}
}
\caption{Raw data from the three experiments conducted on Amazon Mechanical Turk.}
\label{fig:plots_mturk}
\end{figure*}

Let us first eyeball the raw data. Figure~\ref{fig:plots_mturk} presents combined results from the three experiments.  Figure~\ref{fig:plots_breakup} shows the breakup of the evaluations of all the collected responses. The magnitude of the evaluation represents the number of options selected and its sign denotes whether the correct option was selected (positive) or not (negative). Figure~\ref{fig:plots_nonskip} depicts the fraction of responses to attempted questions that turned out to be wrong. Figure~\ref{fig:plots_one} depicts the fraction of responses that were correct when only one option was selected.  Figure~\ref{fig:plots_payment} depicts the average payment per worker. Using this data, let us now investigate the three questions posed at the beginning of the experiments:

(i) Are the workers making a judicious use of the approval voting setup? One can observe from Figure~\ref{fig:plots_breakup} more than $40\%$ responses comprised a selection of two or three options, suggesting that the workers did understand the concept of approval voting.

(ii) Does the presence of a mechanism makes a difference? We compared the data from the approval voting setup under the fixed mechanism with the data from the approval voting setup under Mechanism~\ref{algo:incentive_minbelief}. In particular, we applied Hotelling's T-squared test, where we treated the response by any worker to any question as a two-dimensional data point, with the number options selected and the correctness of the answer as the two dimensions. The results of this test are listed in Table~\ref{tab:HotellingFixedVsOur}. We could reject the null hypothesis (of the two sets of data being drawn from distributions with identical means) with $p < 0.01$ for each of the three experiments.

\begin{table}
\newcommand{\rowHotelling}[4]{#1 & #2 & #3 & #4}
\centering
\begin{tabular}{|l|ccc|}
\hline
\rowHotelling{Experiment}{$T^2$}{$F$}{$p$}\\
\hline
\rowHotelling{Languages}{15.7}{7.8}{0.0004}\\
\rowHotelling{Textures}{21.3}{10.7}{0.000025}\\
\rowHotelling{Animals}{10.2}{5.1}{0.0062}\\
\hline
\end{tabular}
\caption{Hotelling's T-squared test comparing the data from the fixed payment mechanism and the data from Mechanism~\ref{algo:incentive_minbelief} proposed in this paper.}
\label{tab:HotellingFixedVsOur}
\end{table}

(iii) Is there is an opposition from the workers to the interface or the mechanism for any reason? We also elicited feedback about the task from every worker, informing them that the feedback will not affect their payment. We received mostly neutral feedback, some positive feedback, and no negative feedback about either the approval voting interface or our mechanism. 

All in all, these preliminary experiments indicate that our mechanism is practical and can potentially be useful for many applications in machine learning, paying higher amounts to good workers and lower amounts to freeloaders or spammers.

A concluding remark. A standard means of denoising data from crowdsourcing is to ask every question to multiple workers, and employ a statistical aggregation algorithm to aggregate the data so obtained. In the future, we wish to evaluate the performance of our proposed interface and mechanism on such aggregated data. To this end, our goal for the future is to design algorithms designed towards statistical aggregation of data collected through the interface and mechanism proposed in this paper.  

\section{General utility functions}\label{sec:utility}
In this section, we consider a setting where the worker, instead of maximizing her expected payment, aims to maximize the expected value of some \textit{utility function} of her payment.

Consider any function $\utilityfn: \mathbb{R} \rightarrow \mathbb{R}$. Suppose that instead of aiming to maximize the expected payment, the worker has some \textit{utility} $\utilityfn$ for any payment made to her, and that she aims to maximize the expected utility. In other words, for any payment $f$ made to the worker (based on the evaluation of her answers to the gold standard questions), her utility for this payment is $\utilityfn(f)$. The worker aims to maximize the expected value of $\utilityfn(f)$. 

We will require the function $\utilityfn$ to be strictly increasing and invertible. The results presented so far in the paper implicitly assumed that the utility is simply the identity function, namely $\utilityfn(x) = x$. The function $\utilityfn$ is assumed to be public knowledge.

Given the evaluations $\ans{1},\ldots,\ans{\numgold}$ of the worker's responses to the $\numgold$ gold standard questions, consider the following payment mechanism:
\begin{align}
f(\ans{1},\ldots,\ans{\numgold}) = U^{-1}\left( (U(\maxpay)-U(\minpay)) (1-\minbelief)^{\sum_{i = 1}^G (\ans{i}-1)} \prod_{i = 1}^G\mathbf{1}\{\ans{i} \geq 1\} + U(\minpay)\right).
\label{eq:mechanism_utility}
\end{align}

It is easy to see that the properties of Mechanism~\ref{algo:incentive_minbelief} carry over to this mechanism in the case of a general utility function $\utilityfn$. This feature is formalized in the following proposition.
\begin{proposition}\label{thm:utility}
For a worker who aims to maximize function $\utilityfn$ of the payment, the mechanism in Equation~\eqref{eq:mechanism_utility} is incentive compatible, frugal, and is the one and only incentive compatible mechanism to satisfy the no-free-lunch axiom.
\end{proposition}

\section{An alternative problem statement}\label{sec:thresholdEvery}
In earlier sections, we made the coarse belief assumption of the existence of some $\minbelief \in (0,\frac{1}{\numchoices})$ such that the belief of a worker for any option is assumed to either equal $0$ or more than $\minbelief$. We then designed a mechanism to elicit the support of the worker's belief under this assumption. A natural question that arises is that instead of making a coarse belief assumption, can we fix a parameter, say, $\thresholdEvery \in (0,1)$, and incentivize the worker to select all options for which her belief is strictly greater $\thresholdEvery$? Although not the primary focus of this paper, we devote the present section to investigating this complimentary setting out of intellectual curiosity as well as practical relevance. As we show below, the answer to this question is both yes and no.

\subsection{Problem setting}
For a given value of $\thresholdEvery \in (0,1)$, we will call a mechanism as incentive compatible if the expected payment of any worker is strictly maximized when the worker selects all options for which her belief is strictly greater than $\thresholdEvery$. 

We retain most notation form Section~\ref{sec:problem}, with a few exceptions as follows. We continue to let $\paysymbol$ denote the payment function; $\paysymbol: \{-(\numchoices-1),\ldots,\numchoices\}^\numgold \rightarrow [\minpay, \maxpay]$. Observe that unlike the setting considered earlier in Section~\ref{sec:problem}, here we have included $0$ in the domain of the payment function. This is because under the present setting, when $\thresholdEvery \geq \frac{1}{\numchoices}$, there is a possibility that the worker has a belief no more than $\thresholdEvery$ for each option, for instance, if the worker is totally clueless.

Let us define two integers $\selmin$ and $\selmax$ as $\selmin = \indicator{\thresholdEvery < \frac{1}{\numchoices} }$ and $\selmax = \min\{ \lceil \frac{1}{\thresholdEvery} \rceil - 1  , \numchoices \}$.\footnote{The function $\mathbf{1}:\{True,False\}\rightarrow \{0,1\}$ is the indicator function, with $\indicator{x} = 1$ if $x$ is true, and $0$ otherwise.} Observe that if if $\thresholdEvery < \frac{1}{\numchoices}$ then it is meaningless to let the worker select zero options since the belief for at least one option must be $\frac{1}{\numchoices}$ or higher. Also observe that for any value of $\thresholdEvery \in (0,1)$, it is meaningless to allow the worker to select $\lceil \frac{1}{\thresholdEvery} \rceil$ or more options, since it is mathematically impossible for those many options to have beliefs more than $\thresholdEvery$. As a result, we will require the worker to select at least $\selmin$ and at most $\selmax$ options for any question. The goal remains to design the payment function $\pay{\ans{1},\ldots,\ans{\numgold}}$ when $\absolute{\ans{i}} \in \{ \selmin, \ldots, \selmax\}$ for every $i \in [\numgold]$. If the worker's responses do not satisfy this condition, then we assume the convention of setting the payment to a small enough value (say, $\minpay$ or some further penalty). 

We do \emph{not} assume the restriction of coarseness of the beliefs. We stick to the identity utility function, while noting that extension to other utility functions is straightforward following Section~\ref{sec:utility}. 

Finally, we note some special cases which we exclude from the subsequent analysis. The case of $\thresholdEvery = 0$ degenerates to the impossibility result of Theorem~\ref{thm:support_impossible} proved earlier. The cases of $\numchoices=2$ \emph{or} $\thresholdEvery \geq \frac{1}{2}$ degenerate to the ``skip-based'' single-selection setting studied in~\citett{shah2015double}. Hence we focus on the case of $\numchoices \geq 3$ and $\thresholdEvery \in (0, \frac{1}{2} )$ in the rest of this section.

\subsection{Mechanism}

\begin{algorithm}[h!]
 \floatname{algorithm}{Mechanism}
\begin{itemize}[leftmargin=*]
\item \textbf{Input:} Evaluations of the worker's answers to the $\numgold$ gold standard questions $(\ans{1},\ldots,\ans{\numgold})$
\item \textbf{Output:} Define a function $g:\mathbb{R} \rightarrow \mathbb{R}$ as
\begin{align*}
g(y) = (\numchoices - |y|)\thresholdEvery + \mathbf{1}\{ y \geq 1 \}.
\end{align*}
The worker's payment is
\begin{equation*}
f(\ans{1},\ldots,\ans{\numgold}) = a + b\sum_{i=1}^{\numgold} g(\ans{i}),
\end{equation*}
where $a = \minpay$ and $b = \frac{\maxpay - \minpay}{\numgold((\numchoices-1)\thresholdEvery+1)}$.
\end{itemize}
\caption{Incentive mechanism for the alternative problem formulation}
\label{algo:incentive_every}
\end{algorithm}

Given the conventions described in the previous subsection for the payment function, it remains to construct the payment function under ``normal'' conditions, that is, when $\selmin \leq \ans{i} \leq \selmax$ for every $i \in [\numgold]$. Mechanism~\ref{algo:incentive_every} now presents our proposed mechanism for this setting. 

\begin{theorem}\label{thm:possible_everyoption}
Consider any $\thresholdEvery \in (0, \frac{1}{2})$, $\numques \geq \numgold \geq 1$ and $\numchoices \geq 3$. Consider the goal of designing a mechanism such that for each question, the worker is incentivized to select every option for which her belief is more than $\thresholdEvery$. Assume that no belief equals exactly $\thresholdEvery$. Then Mechanism~\ref{algo:incentive_every} is incentive compatible.
\end{theorem}
The function $g$, in words, penalizes the selection of an incorrect option by $\thresholdEvery$ and rewards the selection of the correct option by $1$. Under beliefs $\{\probopt{1}{},\ldots,\probopt{\numchoices}{}\}$ for a gold standard question, when the worker answers as per our requirements, the expected value of $g$ equals $\sum_{i=1}^{\numchoices} \max\{\probopt{i}{},\thresholdEvery\}$. 

The setting also permits a ``multiplicative'' mechanism, consistent with the earlier results in this paper.
\begin{corollary}\label{cor:multiplicative_everyoption}
Under the assumption that no belief equals exactly $\thresholdEvery$, the mechanism \begin{align*}f(\ans{1},\ldots,\ans{\numgold}) = a + b\prod_{i=1}^{\numgold} (g(\ans{i}) - c),\end{align*} for some constants $a$, $b>0$ and $c \leq g(-\selmax)$, is also incentive compatible.
\end{corollary}

\subsection{Uniqueness and an impossibility result}

In this section, we show that the core structure of Mechanism~\ref{algo:incentive_every}, namely the function $g$, is essential for any mechanism. We also show that the (mild) assumption of no belief equalling exactly $\thresholdEvery$ is unavoidable.

\begin{theorem}\label{thm:impossible_everyoption}
Consider any $\thresholdEvery \in (0, \frac{1}{2})$ and any $\numchoices \geq 3$. Consider the goal of designing a mechanism such that for each question, the worker is incentivized to select every option for which her belief is more than $\thresholdEvery$. Then:\\
{\bf (A)} Under the assumption that no belief equals exactly $\thresholdEvery$, when $\numgold = 1$, the function $g$ is the one and only incentive-compatible mechanism upto a constant shift and positive scaling.\\
{\bf (B)} For any $\numques \geq \numgold \geq 1$, no mechanism is incentive compatible in the absence of this assumption.
\end{theorem}

While we do not have a complete answer as to what the ``best'' or ``unique'' mechanism is for general values of $\numques$ and $\numgold$, but going by results proved earlier in the paper, we conjecture that the multiplicative version of the mechanism (Corollary~\ref{cor:multiplicative_everyoption}) may possess attractive properties. Further exploration of this setting is beyond the scope of this paper.


\section{Discussion and open problems}\label{sec:conclusion}
Our goal is to deliver high quality labels for machine learning applications, at low costs, by means of incentive mechanisms or aggregation algorithms or both. In this paper, we pursue the former approach. We take an approval-voting based means of gathering labeled data from crowdsourcing. We design an incentive mechanism via a principled theoretical approach, and prove appealing properties of optimality and uniqueness of our proposed mechanism. Preliminary experiments conducted on Amazon Mechanical Turk corroborate the usefulness of this mechanism for practical scenarios. Our mechanism may also draw more experts to the crowdsourcing platform since their compensation will be significantly higher than that of mediocre workers, unlike most compensation mechanisms in current use.

We conclude with a discussion on closely related topics that merit investigation in the future.

{\bf Aggregation of labels.} 
For the traditional single-selection setting, there is a long, existing line of work on statistical methods to aggregate redundant noisy data from multiple workers~\citepp{dawid1979maximum,whitehill2009whose,raykar2010learning,karger2011iterative,liu2012variational,zhou2012learning}. An open problem is the design of aggregation algorithms for  approval-voting-based data: algorithms that can exploit the specific structure of the responses that arise as a result of the approval voting interface and the proposed mechanism. There is indeed work on aggregation algorithms~\citepp{masso2008weighted,caragiannis2010approximation,brams2014satisfaction,procaccia2015approval} and  probabilistic models~\citepp{marley1993aggregation,falmagne1996random,doignon2004repeated,regenwetter2004approval} for approval-voting in the context of social choice theory; their objective, however, is primarily of fairness and stretgyproofing of the voting procedure, as opposed to our goal of denoising data obtained from multiple heterogeneous workers as required for labeling tasks in crowdsourcing.

{\bf Choosing the right interface.} There are tradeoffs between various interfaces for crowdsourcing. For instance, the approval voting interface elicits the support of the belief whereas the single selection interface elicits the mode. Choosing among these two interfaces would depend on the application under consideration, and moreover, one may adaptively switch between the two depending on the data obtained. A natural question that one may further ask is, why not elicit the entire belief distribution itself? While the entire belief distribution seems to supercede the support and the mode, stating the distribution will also require much more time and effort from the workers, and often also suffer from a higher noise. These tradeoffs must be taken into account when choosing the interface for the application at hand.

{\bf The coarse beliefs parameter.} One may wish to evaluate the value of $\minbelief$ by explicitly asking workers on the crowdsourcing platform for this value. However, it is noted in the literature (e.g., see~\citett{shah2015topology} for experiments on Amazon Mechanical Turk) that the cardinal representations that humans provide are not always consistent with their respective mental beliefs, and are far noisier. This phenomenon suggests the requirement of developing alternative methods of evaluating this parameter. Indeed, measurement is considered one of the most difficult parts of behavioral research.

We look forward to future work exploring these topics in depth.


\section*{Acknowledgements}
This paper was presented in part at the International Conference on Machine Learning (ICML) 2015. The work of the first author was supported in part by a Microsoft Research PhD fellowship.

\appendix

{\vspace{1cm} \Large \bf \noindent APPENDIX}

\section{Proofs}
In this section, we present proofs of the various theoretical results presented in the paper.

\subsection{Proof of Theorem~\ref{thm:support_impossible}: Impossibility}
We assume that there indeed exists some incentive-compatible payment function $\paysymbol$, and prove a contradiction.

Let us first consider the special case of $\numques = \numgold = 1$ and $\numchoices = 2$. Since $\numques = \numgold = 1$, there is only one question. Let $\probopt{1}{} > 0.5$ be the probability, according to the belief of the worker, that option $1$ is correct; the worker then believes that option $2$ is correct with probability $(1-\probopt{1}{})$.

When $\probopt{1}{} = 1$, we need the worker to select option $1$ alone. Thus we need
\begin{align*}
\pay{1} > \pay{2}.
\end{align*}

When $\probopt{1}{} \in (0.5,1)$, we require the worker to select options $1$ and $2$, as opposed to selecting option $1$ alone. For this we need
\begin{align*}
p_1 \pay{1} + (1-p_1) \pay{-1} < \pay{2}
\end{align*}
It follows that we need
\begin{align}
(1-\probopt{1}{})(\pay{1} - \pay{-1}) > \pay{1}-\pay{2}.
\label{eq:support_impossible_1}
\end{align}
However, the inequality~\eqref{eq:support_impossible_1} is satisfied only when $\pay{1} > \pay{-1}$ and $(1-\probopt{1}{}) > \frac{\pay{1}-\pay{2}}{\pay{1} - \pay{-1}}$. Thus for any given payment function $\paysymbol$, a worker with belief $(1-\probopt{1}{}) \in (0, \frac{\pay{1}-\pay{2}}{\pay{1} - \pay{-1}})$ will not be incentivized to select the support of her belief. This yields a contradiction.

We now move on to the general case of $\numques \geq \numgold \geq 1$ and $\numchoices \geq 2$. Consider a worker who is clueless about questions $2$ through $\numques$ (i.e., her belief is uniform across all options for these questions). Suppose this worker selects all $\numchoices$ options for these questions as desired. For the first question, suppose that the worker is sure that options $3,\ldots,\numchoices$ are incorrect. We are now left with the first question and the first two options for this question. Letting $X$ denote a random variable representing the evaluation of the worker's response to the first question, the expected payment then is
\begin{align*}
\frac{\numgold}{\numques} \expect{ \pay{X, \numchoices,\ldots,\numchoices} } + (1-\frac{\numgold}{\numques}) \pay{\numchoices,\ldots,\numchoices}.
\end{align*} 
The expectation in the first term is taken with respect to the randomness in $X$. Defining 
\begin{align*}
\tilde{\paysymbol}(X) \defn \frac{\numgold}{\numques} \pay{X, \numchoices,\ldots,\numchoices} + (1-\frac{\numgold}{\numques}) \pay{\numchoices,\ldots,\numchoices},
\end{align*} 
and applying the same arguments to $\tilde{\paysymbol}$ as those for $\paysymbol$ for the case of $\numques=\numgold=1,\,\numchoices=2$ above gives the desired contradiction. This thus completes the proof of impossibility.

\subsection{Proof of Lemma~\ref{lem:continuity_minbelief}: The workhorse lemma}
Consider some $\minbelief_0 \in (\minbelief, \frac{1}{\numchoices})$. 
Consider a worker such that for every question $i \in \mathcal{I}$, her belief is $\minbelief_0$ for the first option and $\frac{1-\minbelief_0}{\attempt{i}-1}$ for each of the last $(\attempt{i}-1)$ options. For every question $i \notin \mathcal{I}$, her belief is uniformly distributed among the first $\attempt{i}$ options. Now, if the worker selects precisely the support of her beliefs for every question then her expected payment $\Epay_1$ is
\begin{align}
\Epay_1 = 
\frac{1}{{\numques \choose \numgold}} \sum_{(j_1,\ldots,j_\numgold) \subseteq [\numques]} \pay{\attempt{j_1},\ldots,\attempt{j_\numgold}}.
\label{eq:epay1}
\end{align}
We will compare the aforementioned action to another action, where for each question $i \in \mathcal{I}$, the worker selects only the last $(\attempt{i}-1)$ options but not the first option; for each question $i \notin \mathcal{I}$, the worker selects the support of her belief. Under this action, the expected payment $\Epay_2$ is
\begin{align}
\Epay_2 = 
\frac{1}{{\numques \choose \numgold}} \sum_{\substack{(j_1,\ldots,j_\numgold)\\ \subseteq [\numques]}} \sum_{\substack{(\epsilon_1,\ldots,\epsilon_\numgold ) \\ \in \{-1,1\}^\numgold }} \indicator{\{j_i \suchthat \epsilon_i=-1\} \subseteq \mathcal{I}}  (1-\minbelief_0)^{\cardinality{\mathcal{I} \cap \{j_i \suchthat \epsilon_i=1\}}} \minbelief_0^{\cardinality{\mathcal{I} \cap  \{j_i \suchthat \epsilon_i=-1\}}}  \pay{\epsilon_1 \attempt{j_1}',\ldots, \epsilon_\numgold \attempt{j_\numgold}'}.
\label{eq:epay2_complete}
\end{align}
In the expression~\eqref{eq:epay2_complete}, the outer summation represents the expectation over the random choice of the $\numgold$ gold standard questions among the $\numques$ questions. The inner summation represents the expectation with respect to the correctness or incorrectness of the answers to the $\numgold$ gold standard questions: for any question $i$, $\epsilon_i=1$ captures the event where the $i^{\rm th}$ question in the gold standard is answered correctly and $\epsilon_i=-1$ represents the event of this question being answered incorrectly. The term $\indicator{\{j_i \suchthat \epsilon_i=-1\} \subseteq \mathcal{I}}$ ensures that only the questions in $\mathcal{I}$ can be wrong, since it is only these questions for which the worker has selected a subset of her belief's support. 

Since $\pay{x} \geq 0$ for all $x$, we can lower bound $\Epay_2$ as
\begin{align}
\Epay_2 \geq 
\frac{1}{{\numques \choose \numgold}} \sum_{(j_1,\ldots,j_\numgold) \subseteq [\numques]} (1-\minbelief_0)^{\cardinality{\mathcal{I} \cap \{j_1,\ldots, j_\numgold\} } } \pay{\attempt{j_1}',\ldots,\attempt{j_\numgold}'}.
\label{eq:epay2}
\end{align}
An incentive compatible mechanism must incentivize the worker to perform the first action (over the second), i.e, must have $\Epay_1 > \Epay_2$. Thus from~\eqref{eq:epay1} and~\eqref{eq:epay2}, we get
\begin{align}
\frac{1}{{\numques \choose \numgold}} \sum_{(j_1,\ldots,j_\numgold) \subseteq [\numques]} \pay{\attempt{j_1},\ldots,\attempt{j_\numgold}} > \frac{1}{{\numques \choose \numgold}} \sum_{(j_1,\ldots,j_\numgold) \subseteq [\numques]} (1-\minbelief_0)^{\cardinality{\mathcal{I} \cap \{j_1,\ldots, j_\numgold\} } } \pay{\attempt{j_1}',\ldots,\attempt{j_\numgold}'}.
\label{eq:epay_inequality}
\end{align}
Note that~\eqref{eq:epay_inequality} must hold for all $\minbelief_0 > \minbelief$. The left hand side of~\eqref{eq:epay_inequality} does not involve $\minbelief_0$ whereas the right hand side is continuous in $\minbelief_0$. It follows that
\begin{align}
\frac{1}{{\numques \choose \numgold}} \sum_{(j_1,\ldots,j_\numgold) \subseteq [\numques]} \pay{\attempt{j_1},\ldots,\attempt{j_\numgold}} \geq \frac{1}{{\numques \choose \numgold}} \sum_{(j_1,\ldots,j_\numgold) \subseteq [\numques]} (1-\minbelief)^{\cardinality{\mathcal{I} \cap \{j_1,\ldots, j_\numgold\} } } \pay{\attempt{j_1}',\ldots,\attempt{j_\numgold}'}.
\label{eq:epay3}
\end{align}
This proves the first part of the lemma.

We now move on to the second part of the lemma, concerning equality in~\eqref{eq:epay3}. Suppose $\pay{\epsilon_1 \attempt{j_1}',\ldots, \epsilon_\numgold \attempt{j_\numgold}'}$ is strictly positive for any $(j_1,\ldots,j_\numgold) \subseteq [\numques],~\{(\epsilon_1,\ldots,\epsilon_\numgold) \in \{-1,1\}^{\numgold} \backslash \{1\}^\numgold  \suchthat \epsilon_i = 1 ~\mbox{whenever}~ j_i \notin \mathcal{I} \}$. Then~\eqref{eq:epay3} will necessarily be a strict inequality. The claimed necessary condition for equality is thus established.

\subsection{Proof of Theorem~\ref{thm:unique}: Frugality}
\newcommand{\indDistinct}{\gamma} 
\newcommand{\indStep}{\sigma} 
\newcommand{\indVal}{\beta} 
Without loss of generality, assume that $\minpay = 0$ since in our setting, the property of incentive compatibility is invariant to any constant shift and positive scale of the payment. We adopt the succinct notation of $\maxpayproofs \defn \maxpay - \minpay$.

Consider any incentive compatible mechanism $\paysymbol$ such that $\pay{1,\ldots,1} = \maxpayproofs$ and $\pay{\numchoices,\ldots,\numchoices} = (1-\minbelief)^{\numgold(\numchoices-1)} \maxpayproofs$. We will show that this payment mechanism must be identical to Mechanism~\ref{algo:incentive_minbelief}.

We consider the set of evaluations $\anssym$ whose elements are non-decreasing, i.e., $\ans{1} \geq \ans{2} \geq \cdots \geq \ans{\numgold}$; The proof for any other ordering follows in an identical manner.

First consider any $\anssym$ such that $\anssym_\numgold > 0$.
\begin{itemize}
\item Let $\indDistinct(\anssym)$ denote the number of distinct entries in $\anssym$:
\begin{align*}
\indDistinct(\anssym) \defn 1 + \sum_{i=1}^{\numgold-1} \indicator{\ans{i} \neq \ans{i+1}} 
\end{align*}
\item Let $\indStep(\anssym)$ denote the size of the last jump in $\anssym$:
\begin{align*}
\indStep(\anssym) \defn \ans{j} - \ans{j+1} \qquad \mbox{where~} j = \argmax_{i \in [\numgold-1]} \ans{i} \neq \ans{i+1}
\end{align*}
\item Let $\indVal(\anssym)$ denote the numeric value of $\anssym$ in a $\numchoices$-ary number system:
\begin{align*}
\indVal(\anssym) \defn \sum_{i=1}^{\numgold} \numchoices^{\numgold-i} (\ans{i}-1).
\end{align*} 
\end{itemize}
For example, if $\numchoices=5$, $\numgold=5$ and $\anssym = (5,5,4,1,1)$ then $\indDistinct(\anssym) = |\{5,4,1\}|=3$, $\indStep(\anssym) = 4-1=3$ (where $j=3$), and $\indVal(\anssym) = 4 \cdot 5^4 + 4 \cdot 5^3 + 3 \cdot 5^2 + 0 \cdot 5^1 + 0 \cdot 5^0 = 3075$.
The proof involves three nested levels of induction: on $\indDistinct$, on $\indStep$ and then on $\indVal$.

We first induct on $\indDistinct$. The base case is the set $\{\anssym| \indDistinct(\anssym) = 1 \}$, i.e., the set of vectors which have the same value for all its components. Consider any $\ans{0} \in [\numchoices-1]$. Applying Lemma~\ref{lem:continuity_minbelief} with $\attemptsym = (\ans{0}+1,\ldots,\ans{0}+1)$ and $\attemptsym' = (\ans{0},\ldots,\ans{0})$ gives
\begin{align*}
\pay{\ans{0}+1,\ldots,\ans{0}+1} \geq (1-\minbelief)^\numgold \pay{\ans{0},\ldots,\ans{0}}.
\end{align*}
Since this inequality is true for every $\ans{0} \in [\numchoices-1]$, we have
\begin{align*}
\pay{\numchoices,\ldots,\numchoices} \geq (1-\minbelief)^{(\numchoices-\ans{0}) \numgold} \pay{\ans{0},\ldots,\ans{0}} \geq (1-\minbelief)^{(\numchoices-1)\numgold} \pay{1,\ldots,1}.
\end{align*}
Setting $\pay{\numchoices,\ldots,\numchoices}  = (1-\minbelief)^{(\numchoices-1)} \maxpayproofs$ and $\pay{1,\ldots,1} = \maxpayproofs$ proves the base case.

Now suppose our hypothesis is true for all $\{\anssym | \indDistinct(\anssym) \leq \indDistinct_0-1\}$ for some $\indDistinct_0 \in \{2,\ldots,\numchoices\}$. We will now prove that the hypothesis is also true for all $\{\anssym | \indDistinct(\anssym) \leq \indDistinct_0\}$. Towards this goal, we will now induct on $\indStep$. The set of all $\{\anssym | \indDistinct(\anssym) = \indDistinct_0-1\}$ can be treated as a base case for our induction, with this base case corresponding to $\indStep=0$. Due to the induction hypothesis on $\indDistinct$, the base case of $\indStep=0$ is already proven.

Now suppose that the hypothesis is true for all $\{\anssym | \indDistinct(\anssym) = \indDistinct_0,~\indStep(\anssym) \leq \indStep_0-1\}$ for some $\indStep_0 \in [\numchoices-1]$. We will prove that the hypothesis remains true for all $\{\anssym | \indDistinct(\anssym) = \indDistinct_0,~\indStep(\anssym) = \indStep_0\}$. To this end, we will induct on $\indVal$.

Recall that we have restricted our attention to those $\anssym$ which have their elements in a descending order. Observe that the element with the minimum value of $\indVal$ in the set $\{\anssym | \indDistinct(\anssym) = \indDistinct_0,~\indStep(\anssym) = \indStep_0\}$ is $(\indDistinct_0+\indStep_0-1,\ldots,\indStep_0+1,1,\ldots,1)$. We will prove the hypothesis for this element as the base case for our induction on $\indVal$. Applying Lemma~\ref{lem:continuity_minbelief} with $\attemptsym = (\indDistinct_0+\indStep_0-1,\ldots,\indStep_0+2, \indStep_0+1,1,\ldots,1)$ and $\attemptsym' = (\indDistinct_0+\indStep_0-1,\ldots,\indStep_0+2, \indStep_0,1,\ldots,1)$ gives the inequality
\begin{align}
 \const_1 & \pay{\indDistinct_0+\indStep_0-1,\ldots,\indStep_0+2, \indStep_0+1, 1, \ldots, 1} + \const_1' \pay{\indDistinct_0+\indStep_0-1,\ldots,\indStep_0+2, 1, 1, \ldots, 1} \nonumber \\
& + \sum_{s \subsetneq	\{\indDistinct_0+\indStep_0-1,\ldots,\indStep_0+2\} } \left( \const_s \pay{s,1,1,\ldots,1} + \const_s' \pay{s, \indStep_0+1, 1, \ldots, 1} \right) \nonumber \\ 
\geq  \const_1 & (1-\minbelief) \pay{\indDistinct_0+\indStep_0-1,\ldots,\indStep_0+2, \indStep_0, 1, \ldots, 1} + \const_1' \pay{\indDistinct_0+\indStep_0-1,\ldots,\indStep_0+2, 1, 1, \ldots, 1} \nonumber \\
&+ \sum_{s \subsetneq	\{\indDistinct_0+\indStep_0-1,\ldots,\indStep_0+2\}} \left( \const_s \pay{s,1,1,\ldots,1} + \const_s' (1-\minbelief) \pay{s, \indStep_0, 1, \ldots, 1} \right),
\label{eq:unique1}
\end{align}
for some positive constants $\const_1,\,\const_1',\,\const_s,\,\const_s'$ (which represent the probabilities of the respective set of $\numgold$ questions being chosen as the $\numgold$ gold standard questions). Now, for any $s \subsetneq	\{\indDistinct_0+\indStep_0-1,\ldots,\indStep_0+2\}$, observe that $\indDistinct(s,\indStep_0+1, 1, \ldots, 1) \leq \indDistinct_0-1$ and $\indDistinct(s,\indStep_0, 1, \ldots, 1) \leq \indStep_0-1$. Thus from our induction hypothesis, we have 
\begin{align}
\pay{s,\indStep_0+1, 1, \ldots, 1} = (1-\minbelief) \pay{s,\indStep_0, 1, \ldots, 1}.
\label{eq:unique_int1}
\end{align} 
Also, $\indDistinct(\indDistinct_0+\indStep_0-1,\ldots,\indStep_0+2, \indStep_0, 1, \ldots, 1) = \indDistinct_0$ and $\indStep(\indDistinct_0+\indStep_0-1,\ldots,\indStep_0+2, \indStep_0, 1, \ldots, 1) = \indStep_0-1$. Consequently from our induction hypothesis, we have 
\begin{align}
\pay{\indDistinct_0+\indStep_0-1,\ldots,\indStep_0+2, \indStep_0, 1, \ldots, 1} = (1-\minbelief)^{\indDistinct_0+\indStep_0-2 + \cdots + \indStep_0+1 + \indStep_0-1} \maxpayproofs.
\label{eq:unique_int2}
\end{align} 
Substituting~\eqref{eq:unique_int1} and~\eqref{eq:unique_int2} in~\eqref{eq:unique1} and canceling out common terms gives
\begin{align*}
\pay{\indDistinct_0+&\indStep_0-1,\ldots,\indStep_0+2, \indStep_0+1, 1, \ldots, 1} \geq (1-\minbelief)^{\indDistinct_0+\indStep_0-2 + \cdots + \indStep_0} \maxpayproofs.
\end{align*}

We will now derive a matching upper bound on $\pay{\indDistinct_0+\indStep_0-1,\ldots,\indStep_0+2, \indStep_0+1, 1, \ldots, 1}$. Applying Lemma~\ref{lem:continuity_minbelief} with $\attemptsym = (\indDistinct_0+\indStep_0-1,\ldots,\indStep_0+1,2,\ldots,2)$ and $\attemptsym' = (\indDistinct_0+\indStep_0-1,\ldots,\indStep_0+1,1,\ldots,1)$ gives
\begin{align}
\const_1 & \pay{\indDistinct_0+\indStep_0-1,\ldots, \indStep_0+1, 2, \ldots, 2} + \sum_{s \subsetneq \{\indDistinct_0+\indStep_0-1,\ldots,\indStep_0+1\} }  \const_s \pay{s,2,\ldots,2} \nonumber \\
\geq \const_1 & (1-\minbelief)^{\numgold - \indDistinct + 1} \pay{\indDistinct_0+\indStep_0-1,\ldots, \indStep_0+1, 1, \ldots, 1} + \sum_{s \subsetneq \{\indDistinct_0+\indStep_0-1,\ldots,\indStep_0+1\} }  \const_s (1-\minbelief)^{\numgold-|s|} \pay{s,1,\ldots,1},
\label{eq:unique2}
\end{align}
for some positive constants $\const_1,\,\const_s$. Now, for any $s \subsetneq	\{\indDistinct_0+\indStep_0-1,\ldots,\indStep_0+2\}$, observe that $\indDistinct(s, 2, \ldots, 2) \leq \indDistinct_0-1$ and $\indDistinct(s, 1, \ldots, 1) \leq \indStep_0-1$. Thus from our induction hypothesis, we have 
\begin{align}
\pay{s, 2, \ldots, 2} = (1-\minbelief)^{\numgold-|s|} \pay{s, 1, \ldots, 1}.
\label{eq:unique_int3}
\end{align} 
Also, $\indDistinct(\indDistinct_0+\indStep_0-1,\ldots, \indStep_0+1, 2, \ldots, 2) \leq \indDistinct_0$ and $\indStep(\indDistinct_0+\indStep_0-1,\ldots, \indStep_0+1, 2, \ldots, 2) = \indStep_0-1$. Consequently from our induction hypothesis, 
\begin{align}
\pay{\indDistinct_0+\indStep_0-1,\ldots, \indStep_0+1, 2, \ldots, 2} = (1-\minbelief)^{\indDistinct_0+\indStep_0-2 + \ldots + \indStep_0 + \numgold - \indDistinct + 1} \maxpayproofs.
\label{eq:unique_int4}
\end{align} 
Substituting these values in~\eqref{eq:unique2} and canceling out common terms gives
\begin{align*}
\pay{\indDistinct_0+\indStep_0-1,\ldots,\indStep_0+2, \indStep_0+1, 1, \ldots, 1} \leq (1-\minbelief)^{\indDistinct_0+\indStep_0-2 + \cdots + \indStep_0} \maxpayproofs.
\end{align*}
We have thus proved that the hypothesis is true for $\anssym = (\indDistinct_0+\indStep_0-1,\ldots,\indStep_0+2, \indStep_0+1, 1, \ldots, 1)$, the base case for our induction on $\indVal$.

Now consider some $\anssym^*$ such that $\indDistinct(\anssym^*) = \indDistinct_0$, $\indStep(\anssym^*) = \indStep_0$ and $\indVal(\anssym^*) = \indVal_0$, for some $\indVal_0$. Let us denote the components of $\anssym^*$ as $\anssym^* = (\ans{1}^*,\ldots,\ans{m}^*, \underbrace{ \indStep_0 + \ans{\numgold}^*, \ldots, \indStep_0 + \ans{\numgold}^*}_{m_1}, \ans{\numgold}^*,\ldots,\ans{\numgold}^* )$ with $\ans{1}^* \geq \ans{2}^* \geq \cdots \geq \ans{m}^* > \indStep_0 + \ans{\numgold}^*$ for some $m \geq 0,\, m_1 \geq 1,\,  m + m_1 < \numgold$. Suppose the hypothesis is true for all $\{\anssym | \indDistinct(\anssym) = \indDistinct_0,~ \indStep(\anssym) = \indStep_0,~\indVal(\anssym) \leq \indVal_0 - 1\}$.  Applying Lemma~\ref{lem:continuity_minbelief} with $\attemptsym = (\ans{1}^*,\ldots,\ans{m}^*, \underbrace{ \indStep_0 + \ans{\numgold}^*, \ldots, \indStep_0 + \ans{\numgold}^*}_{m_1}, \ans{\numgold}^*,\ldots,\ans{\numgold}^*)$ and $\attemptsym' = (\ans{1}^*,\ldots,\ans{m}^*, \underbrace{ \indStep_0 + \ans{\numgold}^*-1, \ldots, \indStep_0 + \ans{\numgold}^*-1}_{m_1}, \ans{\numgold}^*,\ldots,\ans{\numgold}^* )$ gives the inequality
\begin{align}
&\const_1  \pay{\ans{1}^*, \ldots,\ans{m}^*,  \underbrace{ \indStep_0 + \ans{\numgold}^*, \ldots, \indStep_0 + \ans{\numgold}^* }_{m_1}, \ans{\numgold}^*,\ldots,\ans{\numgold}^*} \nonumber \\
& \qquad\qquad\qquad  + \sum_{s \subsetneq \{\ans{1}^*,\ldots,\ans{m}^*, \underbrace{\scriptstyle \indStep_0 + \ans{\numgold}^*, \ldots, \indStep_0 + \ans{\numgold}^*}_{m_1} \} }  \const_s \pay{s,\ans{\numgold}^*,\ldots, \ans{\numgold}^*}\nonumber \\
& \geq 
\const_1  (1-\minbelief)^{m_1} \pay{\ans{1}^*, \ldots,\ans{m}^*,  \underbrace{ \indStep_0 + \ans{\numgold}^* - 1, \ldots, \indStep_0 + \ans{\numgold}^* - 1 }_{m_1}, \ans{\numgold}^*,\ldots,\ans{\numgold}^*} \nonumber \\
& \qquad\qquad\qquad  + \sum_{s \subsetneq  \{\ans{1}^*,\ldots,\ans{m}^*,  \underbrace{\scriptstyle \indStep_0 + \ans{\numgold}^* - 1, \ldots, \indStep_0 + \ans{\numgold}^* - 1}_{m_1} \} }  \const_s (1-\minbelief)^{\sum_i \indicator{s_i = \indStep_0 + \ans{\numgold}^*-1}} \pay{s,\ans{\numgold}^*,\ldots, \ans{\numgold}^*}, \label{eq:unique3}
\end{align}
for some positive constants $\const_1,\,\const_s$. Observe that 
\begin{align*}
\indDistinct(\ans{1}^*, \ldots,\ans{m}^*, \underbrace{ \indStep_0 + \ans{\numgold}^*-1, \ldots, \indStep_0 + \ans{\numgold}^*-1 }_{m_1}, \ans{\numgold}^*,\ldots,\ans{\numgold}^*) = 
\begin{cases}
\indDistinct_0 - 1 & \quad \mbox{if}~\indStep_0 = 1\\
\indDistinct_0 & \quad \mbox{otherwise},
\end{cases}
\end{align*}
and the induction hypothesis is satisfied in the first case. In the second case, 
\begin{align*}
\indStep(\ans{1}^*, \ldots,\ans{m}^*, \underbrace{ \indStep_0 + \ans{\numgold}^*-1, \ldots, \indStep_0 + \ans{\numgold}^*-1 }_{m_1}, \ans{\numgold}^*,\ldots,\ans{\numgold}^*) = \indStep_0 - 1,
\end{align*} 
and hence the induction hypothesis is satisfied in the second case as well. Thus 
\begin{align}
\pay{\ans{1}^*, \ldots,\ans{m}^*, & \underbrace{ \indStep_0 + \ans{\numgold}^*-1, \ldots, \indStep_0 + \ans{\numgold}^*-1 }_{m_1}, \ans{\numgold}^*,\ldots,\ans{\numgold}^*} \nonumber \\
&  = (1-\minbelief)^{\sum_{i=1}^{m}(\ans{i}^*-1) + m_1(\indStep_0 + \ans{\numgold}^*-2) + (\numgold - m_1 - m)(\ans{\numgold}^*-1)} \maxpayproofs.
\label{eq:unique_int5}
\end{align}
For any for any $s \subsetneq	\{\ans{1}^*,\ldots,\ans{m}^*, \underbrace{ \indStep_0 + \ans{\numgold}^* - 1, \ldots, \indStep_0 + \ans{\numgold}^* - 1}_{m_1} \}$, define $\mathfrak{m}_1(s) \defn \sum_i \indicator{s_i = \indStep_0 + \ans{\numgold}^*-1}$. Observe that if $\mathfrak{m}_1(s) > 0$ then either $\indDistinct((s,\ans{\numgold}^*,\ldots,\ans{\numgold}^*)) \leq \indDistinct_0 - 1$ or $\indStep((s,\ans{\numgold}^*,\ldots,\ans{\numgold}^*)) \leq \indStep_0 - 1$; if $\mathfrak{m}_1(s) = 0$ then $\indDistinct((s,\ans{\numgold}^*,\ldots,\ans{\numgold}^*)) \leq \indDistinct_0 - 1$. For any $s \subsetneq	\{\ans{1}^*,\ldots,\ans{m}^*, \underbrace{ \indStep_0 + \ans{\numgold}^*, \ldots, \indStep_0 + \ans{\numgold}^* }_{m_1} \}$, define $\tilde{\mathfrak{m}}_1(s) \defn \sum_i \indicator{s_i = \indStep_0 + \ans{\numgold}^*}$. Observe that if $\tilde{\mathfrak{m}}_1(s) > 0$ then either $\indDistinct((s,\ans{\numgold}^*,\ldots,\ans{\numgold}^*)) \leq \indDistinct_0 - 1$ or $\indVal((s,\ans{\numgold}^*,\ldots,\ans{\numgold}^*)) \leq \indVal_0 - 1$; if $\tilde{\mathfrak{m}}_1(s) = 0$ then $\indDistinct((s,\ans{\numgold}^*,\ldots,\ans{\numgold}^*)) \leq \indDistinct_0 - 1$. Consequently from our induction hypothesis we have  
\begin{align}
\sum_{s \subsetneq \{\ans{1}^*,\ldots,\ans{m}^*, \underbrace{\scriptstyle \indStep_0 + \ans{\numgold}^*, \ldots, \indStep_0 + \ans{\numgold}^*}_{m_1} \} } & \const_s \pay{s,\ans{\numgold}^*,\ldots, \ans{\numgold}^*}\nonumber \\
& =  \sum_{s \subsetneq  \{\ans{1}^*,\ldots,\ans{m}^*,  \underbrace{\scriptstyle \indStep_0 + \ans{\numgold}^* - 1, \ldots, \indStep_0 + \ans{\numgold}^* - 1}_{m_1} \} }  \const_s (1-\minbelief)^{\sum_i \indicator{s_i = \indStep_0 + \ans{\numgold}^*-1}} \pay{s,\ans{\numgold}^*,\ldots, \ans{\numgold}^*}.
\label{eq:unique_int6}
\end{align}
Substituting~\eqref{eq:unique_int5} and~\eqref{eq:unique_int6} in~\eqref{eq:unique3} and canceling out common terms gives
\begin{align*}
\pay{\ans{1}^*,\ldots,\ans{m}^*, & \underbrace{ \indStep_0 + \ans{\numgold}^*, \ldots, \indStep_0 + \ans{\numgold}^*}_{m_1}, \ans{\numgold}^*,\ldots,\ans{\numgold}^*} \\
& \geq (1-\minbelief)^{m_1} \pay{\ans{1}^*,\ldots,\ans{m}^*, \underbrace{ \indStep_0 + \ans{\numgold}^* - 1, \ldots, \indStep_0 + \ans{\numgold}^* - 1}_{m_1}, \ans{\numgold}^*,\ldots,\ans{\numgold}^*}\\
& = (1-\minbelief)^{\sum_{i=1}^{m}(\ans{i}^*-1) + m_1(\indStep_0 + \ans{\numgold}^*-1) + (\numgold - m_1 - m)(\ans{\numgold}^*-1)} \maxpayproofs.
\end{align*} 
We will now employ Lemma~\ref{lem:continuity_minbelief} again to derive a matching lower bound. Setting $\attemptsym = (\ans{1}^*, \ldots, \ans{m}^*,\allowbreak \underbrace{ \indStep_0 + \ans{\numgold}^*, \ldots, \indStep_0 + \ans{\numgold}^*}_{m_1}, \ans{\numgold}^*+1,\ldots,\ans{\numgold}^*+1)$ and $\attemptsym' = (\ans{1}^*,\ldots,\ans{m}^*, \underbrace{ \indStep_0 + \ans{\numgold}^*, \ldots, \indStep_0 + \ans{\numgold}^*}_{m_1}, \ans{\numgold}^*,\ldots,\ans{\numgold}^*)$ in Lemma~\ref{lem:continuity_minbelief} yields the inequality
\begin{align}
&\const_1  \pay{\ans{1}^*, \ldots,\ans{m}^*,  \underbrace{ \indStep_0+ \ans{\numgold}^*, \ldots, \indStep_0+ \ans{\numgold}^* }_{m_1}, \ans{\numgold}^*+1,\ldots,\ans{\numgold}^*+1} \nonumber \\
& \qquad\qquad\qquad  + \sum_{s \subsetneq \{\ans{1}^*,\ldots,\ans{m}^*, \underbrace{\scriptstyle \indStep_0 + \ans{\numgold}^*, \ldots, \indStep_0 + \ans{\numgold}^*}_{m_1} \} }  \const_s \pay{s,\ans{\numgold}^*+1,\ldots, \ans{\numgold}^*+1}\nonumber \\
& \geq 
\const_1  (1-\minbelief)^{m_1} \pay{\ans{1}^*, \ldots,\ans{m}^*,  \underbrace{ \indStep_0+ \ans{\numgold}^*, \ldots, \indStep_0 + \ans{\numgold}^*}_{m_1}, \ans{\numgold}^*,\ldots,\ans{\numgold}^*} \nonumber \\
& \qquad\qquad\qquad  + \sum_{s \subsetneq  \{\ans{1}^*,\ldots,\ans{m}^*,  \underbrace{\scriptstyle \indStep_0+ \ans{\numgold}^*, \ldots, \indStep_0 + \ans{\numgold}^*}_{m_1} \} }  \const_s (1-\minbelief)^{\numgold - \cardinality{s}} \pay{s,\ans{\numgold}^*,\ldots, \ans{\numgold}^*},
\label{eq:unique4}
\end{align}
for some positive constants $\const_1,\,\const_s$. Observe that 
\begin{align*}
\indDistinct(\ans{1}^*, \ldots,\ans{m}^*, \underbrace{ \indStep_0 + \ans{\numgold}^*, \ldots, \indStep_0 + \ans{\numgold}^*}_{m_1}, \ans{\numgold}^*+1,\ldots,\ans{\numgold}^*+1) = 
\begin{cases}
\indDistinct_0 - 1 & \quad \mbox{if}~\indStep_0 = 1\\
\indDistinct_0 & \quad \mbox{otherwise},
\end{cases}
\end{align*}
and that the induction hypothesis is satisfied in the first case. In the second case, 
\begin{align*}
\indStep(\ans{1}^*, \ldots,\ans{m}^*, \underbrace{ \indStep_0 + \ans{\numgold}^*, \ldots, \indStep_0 + \ans{\numgold}^*}_{m_1}, \ans{\numgold}^*+1,\ldots,\ans{\numgold}^*+1) = \indStep_0 - 1,
\end{align*} 
and hence the induction hypothesis is satisfied in the second case as well. Thus 
\begin{align}
\pay{\ans{1}^*, \ldots,\ans{m}^*, & \underbrace{ \indStep_0 + \ans{\numgold}^*, \ldots, \indStep_0 + \ans{\numgold}^* }_{m_1}, \ans{\numgold}^*+1,\ldots,\ans{\numgold}^*+1} \nonumber\\ 
&= (1-\minbelief)^{\sum_{i=1}^{m}(\ans{i}^*-1) + m_1(\indStep_0 + \ans{\numgold}^*-1) + (\numgold - m_1 - m)(\ans{\numgold}^*-2)} \maxpayproofs.
\label{eq:unique_int7}
\end{align}
Now consider any $s \subsetneq	\{\ans{1}^*,\ldots,\ans{m}^*, \underbrace{ \indStep_0 + \ans{\numgold}^* , \ldots, \indStep_0 + \ans{\numgold}^*}_{m_1} \}$, and recall our notation of $\tilde{\mathfrak{m}}_1(s) \defn \sum_i \indicator{s_i = \indStep_0 + \ans{\numgold}^*}$. If $\indStep_0=1$ or if $\tilde{\mathfrak{m}}_1(s) = 0$ then $\indDistinct((s,\ans{\numgold}^*+1,\ldots, \ans{\numgold}^*+1)) \leq \indDistinct_0 - 1$; if $\indStep > 1$ and $\tilde{\mathfrak{m}}_1(s) > 0$ then $\indDistinct((s,\ans{\numgold}^*+1,\ldots, \ans{\numgold}^*+1)) \leq \indDistinct_0$ and $\indStep(s,\ans{\numgold}^*+1,\ldots, \ans{\numgold}^*+1) \leq \indStep_0-1$. If $\tilde{\mathfrak{m}}_1(s) = 0$ then $\indDistinct((s,\ans{\numgold}^*,\ldots, \ans{\numgold}^*)) \leq  \indDistinct_0 - 1$, otherwise $\indDistinct((s,\ans{\numgold}^*,\ldots, \ans{\numgold}^*)) \leq \indDistinct_0$, $\indStep((s,\ans{\numgold}^*,\ldots, \ans{\numgold}^*)) = \indStep_0$ and $\indVal((s,\ans{\numgold}^*,\ldots, \ans{\numgold}^*)) \leq \indVal_0-1$. These terms thus satisfy our induction hypothesis and hence
\begin{align}
\pay{s,\ans{\numgold}^*+1,\ldots, \ans{\numgold}^*+1} = (1-\minbelief)^{\numgold - \cardinality{s}} \pay{s,\ans{\numgold}^*,\ldots, \ans{\numgold}^*}.
\label{eq:unique_int8}
\end{align}
Substituting~\eqref{eq:unique_int7} and~\eqref{eq:unique_int8} in~\eqref{eq:unique4} gives us our desired matching lower bound
\begin{align*}
\pay{\ans{1}^*,\ldots,\ans{m}^*, & \underbrace{ \indStep_0 + \ans{\numgold}^*, \ldots, \indStep_0 + \ans{\numgold}^*}_{m_1}, \ans{\numgold}^*,\ldots,\ans{\numgold}^*} \leq (1-\minbelief)^{\sum_{i=1}^{m}(\ans{i}^*-1) + m_1(\indStep_0 + \ans{\numgold}^*-1) + (\numgold - m_1 - m)(\ans{\numgold}^*-1)} \maxpayproofs.
\end{align*} 
This completes the proof for $\{\anssym| \ans{i} \geq 0 ~\forall~i \in [\numgold]\}$.

We will now show that $\pay{\anssym} = 0$ for all $\{\anssym \suchthat \min_{i \in [\numgold]} \ans{i} < 0\}$. The arguments above for the case $\{\anssym \suchthat \min_{i \in [\numgold]} \ans{i} > 0\}$ imply that for any incentive-compatible function $\paysymbol$, the first part of Lemma~\ref{lem:continuity_minbelief} must be satisfied with equality. This allows us to employ the second part of Lemma~\ref{lem:continuity_minbelief}. For $i \in [\numgold]$, let $\attempt{i} = \attempt{i}' = \ans{i}$ if $\ans{i} > 0$, and $\attempt{i} - 1 = \attempt{i}' = |\ans{i}|$ otherwise; set $\attempt{i} = \attempt{i}' = \numchoices$ for all $i \in \{\numgold+1,\ldots,\numques\}$. Then the second part of Lemma~\ref{lem:continuity_minbelief} necessitates $\pay{\ans{1},\ldots,\ans{\numgold}}=0$, thus completing the proof.


\subsection{Proof of Theorem~\ref{thm:IC_arbitrary}: Mechanism in absence of coarse belief assumption}
Without loss of generality, assume that $\minpay = 0$ since in our setting, the property of incentive compatibility is invariant to any constant shift and positive scale of the payment. We adopt the succinct notation of $\maxpayproofs \defn \maxpay - \minpay$.

First consider the case of $\numques = \numgold = 1$. Mechanism~\ref{algo:incentive_minbelief} reduces to $\pay{\anssym} = \maxpayproofs (1-\minbelief)^{(\ans{1}-1)} \mathbf{1}\{\ans{1} \geq 0\}$. Suppose without loss of generality that the worker's beliefs for the $\numchoices$ options are $p_1 \geq \cdots \geq p_{\numchoices}$ and suppose $m = \argmax_z \left( \frac{p_{(z)}}{\sum_{i=1}^{z}p_{(i)}} > \minbelief \right)$. A mechanism that is incentive compatible will strictly maximize the worker's expected payment when she selects the options $\{1,\ldots,m\}$. 

Suppose a worker decides to select some $\ell$ of the $\numchoices$ options, say options $\{o_1,\ldots,o_\ell\} \subseteq [\numchoices]$. Then it is easy to see that her expected payment,
\begin{align*}
\maxpayproofs \sum_{i=1}^{\ell} p_{o_i} (1-\minbelief)^{\ell-1},
\end{align*} 
is maximized when she selects options $\{1,\ldots,\ell\}$, i.e., the $\ell$ options that are most likely to be correct. It remains to show that among all choices of $\ell \in [\numchoices]$, the expected payment is maximized when the worker selects $\ell = m$. Let $\Epay_\ell$ denote the expected payment when the worker selects $\ell$ options:
\begin{align*}
\Epay_\ell =  \maxpayproofs \sum_{i=1}^{\ell} p_i (1-\minbelief)^{\ell-1}.
\end{align*}
Hence for any $\ell \in \{2,\ldots,\numchoices\}$, we have
\begin{align*}
\frac{\Epay_{\ell-1}}{\Epay_{\ell}} = \frac{\maxpayproofs \sum_{i=1}^{\ell-1} p_i (1-\minbelief)^{\ell-2}}{\maxpayproofs \sum_{i=1}^{\ell} p_i (1-\minbelief)^{\ell-1}} = \frac{1}{1-\minbelief} \left(1 - \frac{p_\ell}{\sum_{i=1}^{\ell} p_i}\right).
\end{align*}
We know that $\frac{p_\ell}{\sum_{i=1}^{\ell} p_i} < \minbelief$ whenever $\ell > m$, and $\frac{p_\ell}{\sum_{i=1}^{\ell} p_i} > \minbelief$ when $\ell = m$. Furthermore, since $p_\ell$ decreases with $\ell$ and $\sum_{i=1}^{\ell} p_i$ increases with $\ell$, it must also be that $\frac{p_\ell}{\sum_{i=1}^{\ell} p_i} > \minbelief$ for all $\ell < m$. Thus we have $\frac{\Epay_\ell}{\Epay_{\ell-1}} > 1$ for all $\ell \leq m$ and $\frac{\Epay_\ell}{\Epay_{\ell-1}} < 1$ for all $\ell > m$, or in other words,
\begin{align*}
\cdots  < \Epay_{m-2} < \Epay_{m-1} < \Epay_m > \Epay_{m+1} > \Epay_{m+2} > \cdots.
\end{align*} 
It follows that the worker will be incentivized to choose $\ell = m$.

Let us now consider the case of $\numques = \numgold \geq 1$. By our assumption of the independence of the beliefs of the worker across the questions, the expected payment equals
\begin{align*}
\prod_{i=1}^{\numgold} \mathbf{E} \left[ \maxpayproofs (1-\minbelief)^{(\ans{i}-1)} \mathbf{1}\{\ans{i} \geq 0\} \right].
\end{align*}
Since the payments are non-negative, if each individual component in the product is maximized then the product is also necessarily maximized. Each individual component simply corresponds to the setting of $\numques = \numgold = 1$ discussed earlier. Thus calling upon our earlier result, we get that the expected payment for the case $\numques = \numgold \geq 1$ is maximized when the worker acts as desired for every question.

Let us finally consider the general case of $\numques \geq \numgold \geq 1$. Recall from~\eqref{eq:epay_definition} that the expected payment for the general case is a cascade of two expectations: the outer expectation is with respect to the uniformly random distribution of the $\numgold$ gold standard questions among the $\numques$ total questions, while the inner expectation is taken over the worker's beliefs of the different questions conditioned on the choice of the gold standard questions and restricts attention to only these $\numgold$ questions. The arguments above for the case $\numques = \numgold$ prove that every individual term in the inner expectation is maximized when the worker acts as desired. The outer expectation does not affect this argument. The expected payment is thus maximized when the worker acts as desired.

\subsection{Proof of Theorem~\ref{thm:nfl}: Uniqueness}
Without loss of generality, assume that $\minpay = 0$ since in our setting, the property of incentive compatibility is invariant to any constant shift and positive scale of the payment. We adopt the succinct notation of $\maxpayproofs \defn \maxpay - \minpay$. The proof of this theorem employs some of the tools developed in~\citett{shah2015double}. We begin with a lemma deriving a condition that must necessarily be satisfied by any incentive-compatible mechanism. Note that we are not making the coarse belief assumption and supposing that workers can have arbitrary beliefs.

\begin{lemma}
Any incentive-compatible mechanism must satisfy
\begin{align*}
\pay{\ans{1},&\ldots,\ans{i-1},\ans{i}+1,\ans{i+1},\ldots,\ans{\numgold}} \\
& = (1-\minbelief) \pay{\ans{1},\ldots,\ans{i-1}, \ans{i} , \ans{i+1},\ldots,\ans{\numgold}} + \minbelief \pay{\ans{1},\ldots, \ans{i-1},-\ans{i}, \ans{i+1},\ldots, \ans{\numgold}},
\end{align*}
for every $i \in [\numgold]$ and $(\ans{1},\ldots,\ans{i-1},\ans{i+1},\ldots,\ans{\numgold} ) \in \{-(\numchoices-1),\ldots,-1,1,\ldots,\numchoices\}^{{\numgold}-1}$, $\ans{i} \in [\numchoices-1]$.
\label{lem:nec_subset}
\end{lemma}
Note that the lemma does \textit{not} use the no-free-lunch condition. The proof of the lemma is provided at the end of this section. Using this lemma, we now complete the proof of the theorem. 

Consider any incentive-compatible mechanism $\paysymbol$ that satisfies the no-free-lunch condition. 
We first show that the mechanism must necessarily make a zero payment when one more more questions in the gold standard are attempted incorrectly. To this end, observe that since $\paysymbol \geq 0$ and $\minbelief \in (0,1)$, the statement of Lemma~\ref{lem:nec_subset} necessitates that for every $i \in [\numgold]$ and $(\ans{1},\ldots,\ans{i-1},\ans{i+1},\ldots,\ans{\numgold} ) \in \{-(\numchoices-1),\ldots,\numchoices\}^{{\numgold}-1}$, $\ans{i} \in [\numchoices-1]$:
\begin{align*}
\hbox{If~} \pay{&\ans{1},\ldots,\ans{i-1},\ans{i}+1,\ans{i+1},\ldots,\ans{\numgold}} = 0 \\
&\hbox{then~}\pay{\ans{1},\ldots,\ans{i-1}, \ans{i} , \ans{i+1},\ldots,\ans{\numgold}} = \pay{\ans{1},\ldots, \ans{i-1},-\ans{i}, \ans{i+1},\ldots, \ans{\numgold}} = 0.
\end{align*}
A repeated application of this argument implies:
\begin{align*}
\hbox{If~} \pay{\ans{1},\ldots,\ans{i-1},\numchoices,\ans{i+1},\ldots,\ans{\numgold}} = 0 \quad \hbox{then~} \pay{\ans{1},\ldots,\ans{i-1}, \ans{i} , \ans{i+1},\ldots,\ans{\numgold}} = 0,
\end{align*}
for all $\ans{i} \in \{-(\numchoices-1),\ldots,-1,1,\ldots,\numchoices-1\}$.

Now consider any evaluation $(\ans{1},\ldots,\ans{\numgold})$ which has at least one incorrect answer. Suppose without loss of generality that the first question is the one answered incorrectly, i.e., $\ans{1} \leq -1$. The no-free-lunch condition then makes $\pay{\ans{1},\numchoices,\ldots,\numchoices} = 0$. Applying our arguments from above we get that $\pay{\ans{1},\ans{2},\ldots,\ans{\numgold}}=0$ for every value of $(\ans{2},\ldots,\ans{\numgold}) \in \{-(\numchoices-1),\ldots,-1,1,\ldots,\numchoices\}$.

Substituting this necessary condition in Lemma~\ref{lem:nec_subset}, we get that for every question $i \in \{1,\ldots,{\numgold}\}$ and every $(\ans{1},\ldots,\ans{i-1},\ans{i+1},\ldots,\ans{\numgold} ) \in [\numchoices]^{{\numgold}-1}$, $\ans{i} \in [\numchoices-1]$, 
\begin{align*}
\pay{\ans{1},&\ldots,\ans{i-1},\ans{i}+1,\ans{i+1},\ldots,\ans{\numgold}} = (1-\minbelief) \pay{\ans{1},\ldots,\ans{i-1}, \ans{i} , \ans{i+1},\ldots,\ans{\numgold}}.
\end{align*}
Substituting $\pay{1,\ldots,1} = \maxpayproofs$, we get the desired answer.

We now return to complete the proof of Lemma~\ref{lem:nec_subset}.
\begin{proof}[\textbf{Proof of Lemma~\ref{lem:nec_subset}}]
First consider the case of ${\numgold}={\numques}$. Consider some $\eta,\gamma \in \{0,\ldots,{\numgold}-1\}$ with $\eta + \gamma < \numgold$. Suppose $i=\eta+\gamma+1$, $\ans{1},\ldots,\ans{\eta}\in [\numchoices-1]$, $\ans{\eta+1},\ldots,\ans{\eta+\gamma} \in -[\numchoices-1]$ and $\ans{\eta+\gamma+2},\ldots,\ans{\numques}=\numchoices$.

For every question $j \in [\eta + \gamma]$, suppose the worker's belief is $\delta_j \in (0,\minbelief)$ for the last option and $\frac{1-\delta_j}{|\ans{j}|}$ each for the first $|\ans{j}|$ options. One can verify that since $\delta_j < \minbelief < \frac{1}{\numchoices}$ and $|\ans{j}| \leq \numchoices-1$, it must be that $\frac{1-\delta_j}{|\ans{j}|} > \delta_j$, and that incentive-compatibility requires incentivizing the worker to select the first $|\ans{j}|$ options. Suppose the worker does so. Now for every question $j' \in \{\eta + \gamma+2,\ldots,\numques\}$, suppose the belief of the worker is uniform across all $\numchoices$ options. The worker should be incentivized to select all $\numchoices$ options in this case; suppose the worker does so. Finally, for question $i$, suppose the worker's belief is $\delta \in (\frac{\minbelief}{2},\frac{3\minbelief}{2})$ for the last option and $\frac{1-\delta}{|\ans{i}|}$ each for the first $|\ans{i}|$ options. Then the worker must be incentivized to select the first $|\ans{i}|$ options alone if $\delta < \minbelief$, and select the last option along with the first $|\ans{i}|$ options if $\delta > \minbelief$. 

Define $\{r_j\}_{j\in[\eta+\gamma]}$ as $r_j = \delta_j$ for $j\in [\eta]$, and $r_j=1-\delta_j$ for $j \in \{\eta+1,\eta+\gamma\}$. Let $\boldsymbol{\epsilon} :=\{\epsilon_1,\ldots,\epsilon_{\eta+\gamma}\} \in \{-1,1\}^{\eta+\gamma}$.  Incentive-compatibility for question $i$ necessitates
\begin{align*}
&(1-\delta) \sum_{\boldsymbol{\epsilon} \in \{-1,1\}^{\eta+\gamma}} \left(f(\epsilon_1 \ans{1},\ldots,\epsilon_\eta \ans{\eta}, \epsilon_{\eta+1} \ans{\eta+1},\ldots,\epsilon_{\eta+\gamma} \ans{\eta+\gamma}, \ans{i} ,\skipsym,\ldots,\skipsym) \prod_{j \in [\eta+\gamma]} r_j^{\frac{1-\epsilon_{j}}{2}} (1-r_j)^{\frac{1+\epsilon_{j}}{2}}\right)\nonumber\\
&\qquad+
\delta \sum_{\boldsymbol{\epsilon} \in \{-1,1\}^{\eta+\gamma}} \left(f(\epsilon_1\ans{1},\ldots,\epsilon_\eta \ans{\eta},\epsilon_{\eta+1} \ans{\eta+1},\ldots,\epsilon_{\eta+\gamma} \ans{\eta+\gamma}, -\ans{i} ,\skipsym,\ldots,\skipsym) \prod_{j \in [\eta+\gamma]} r_j^{\frac{1-\epsilon_{j}}{2}} (1-r_j)^{\frac{1+\epsilon_{j}}{2}}\right)\nonumber\\
&\overset{\delta>\minbelief}{\underset{\delta < \minbelief}{\lessgtr}}
\sum_{\boldsymbol{\epsilon} \in \{-1,1\}^{\eta+\gamma}} \left(f(\epsilon_1\ans{1},\ldots,\epsilon_\eta \ans{\eta},\epsilon_{\eta+1} \ans{\eta+1},\ldots,\epsilon_{\eta+\gamma} \ans{\eta+\gamma}, \ans{i}+1,\skipsym,\ldots,\skipsym) \prod_{j \in [\eta+\gamma]} r_j^{\frac{1-\epsilon_{j}}{2}} (1-r_j)^{\frac{1+\epsilon_{j}}{2}}\right).
\end{align*}
The left hand side of this expression is the expected payment if the worker chooses the first $|\ans{i}|$ options for question $(\eta+\gamma+1)$, while the right hand side is the expected payment if she chooses the first $|\ans{i}|$ options as well as the last option. For any real-valued variable $q$, and for any real-valued constants $a$, $b$ and $c$, \[a q ~\overset{q<c}{\underset{q>c}{\lessgtr}}~b \quad \Rightarrow \quad \ ac=b~.\] With $q = 1 - \delta$ in this argument, we get
\begin{align}
&(1-\minbelief) \sum_{\boldsymbol{\epsilon} \in \{-1,1\}^{\eta+\gamma}} \left(f(\epsilon_1 \ans{1},\ldots,\epsilon_\eta \ans{\eta}, \epsilon_{\eta+1} \ans{\eta+1},\ldots,\epsilon_{\eta+\gamma} \ans{\eta+\gamma}, \ans{i} ,\skipsym,\ldots,\skipsym) \prod_{j \in [\eta+\gamma]} r_j^{\frac{1-\epsilon_{j}}{2}} (1-r_j)^{\frac{1+\epsilon_{j}}{2}}\right)\nonumber\\
&\qquad+
\minbelief \sum_{\boldsymbol{\epsilon} \in \{-1,1\}^{\eta+\gamma}} \left(f(\epsilon_1\ans{1},\ldots,\epsilon_\eta \ans{\eta},\epsilon_{\eta+1} \ans{\eta+1},\ldots,\epsilon_{\eta+\gamma} \ans{\eta+\gamma}, -\ans{i} ,\skipsym,\ldots,\skipsym) \prod_{j \in [\eta+\gamma]} r_j^{\frac{1-\epsilon_{j}}{2}} (1-r_j)^{\frac{1+\epsilon_{j}}{2}}\right)\nonumber\\
&-
\sum_{\boldsymbol{\epsilon} \in \{-1,1\}^{\eta+\gamma}} \left(f(\epsilon_1\ans{1},\ldots,\epsilon_\eta \ans{\eta},\epsilon_{\eta+1} \ans{\eta+1},\ldots,\epsilon_{\eta+\gamma} \ans{\eta+\gamma}, \ans{i}+1,\skipsym,\ldots,\skipsym) \prod_{j \in [\eta+\gamma]} r_j^{\frac{1-\epsilon_{j}}{2}} (1-r_j)^{\frac{1+\epsilon_{j}}{2}}\right)=0.\label{eq:lineareqn_subset_proof1}
\end{align}
The left hand side of~\eqref{eq:lineareqn_subset_proof1} represents a polynomial in $(\eta+\gamma)$ variables $\{r_j\}_{j=1}^{\eta+\gamma}$ which evaluates to zero for all values of the variables within an $(\eta+\gamma)$-dimensional solid ball. Thus, the coefficients of the monomials in this polynomial must be zero. In particular, the constant term must be zero. The constant term appears when $\epsilon_{j}=1~\forall~j$ in the summations in~\eqref{eq:lineareqn_subset_proof1}. Setting the constant term to zero gives
\begin{align*}
(1-\minbelief) f(\ans{1},\ldots,\ans{\eta+\gamma}, \ans{\eta+\gamma+1} ,\skipsym,\ldots,\skipsym) + \minbelief &f(\ans{1},\ldots,\ans{\eta+\gamma}, -\ans{\eta+\gamma+1},\skipsym,\ldots,\skipsym) \nonumber\\
&- f(\ans{1},\ldots,\ans{\eta+\gamma}, \ans{\eta+\gamma+1}+1,\skipsym,\ldots,\skipsym) = 0
\end{align*}
as desired.
Since the arguments above hold for any permutation of the ${\numques}$ questions, this completes the proof for the case of ${\numgold}={\numques}$.

Now consider the case ${\numgold}<{\numques}$. Let $g:\{-(\numchoices-1),\ldots,-1,1,\cdots,\numchoices\}^{\numques} \rightarrow \mathbb{R}_+$ represent the expected payment given an evaluation of all the ${\numques}$ answers, when the identities of the gold standard questions are unknown. Here, the expectation is with respect to the (uniformly random) choice of the ${\numgold}$ gold standard questions. If $(\ans{1},\ldots,\ans{\numques})\in\{-(\numchoices-1),\ldots,-1,1,\cdots,\numchoices\}^{\numques}$ are the evaluations of the worker's answers to the ${\numques}$ questions then the expected payment is
\begin{align}
g(\ans{1},\ldots,\ans{\numques}) = \frac{1}{{{\numques} \choose {\numgold}}} \sum_{(i_1,\ldots,i_{\numgold})\subseteq\{1,\ldots,{\numques}\}}  \pay{\ans{i_1},\ldots,\ans{i_{\numgold}}}.\label{eq:skip_notation_subsample}
\end{align}
Applying the same arguments to $g$ as done to $\paysymbol$ above, gives
\begin{align}
(1-\minbelief) g(\ans{1},\ldots,\ans{\eta+\gamma}, \ans{\eta+\gamma+1} ,\skipsym,\ldots,\skipsym) + \minbelief & g(\ans{1},\ldots,\ans{\eta+\gamma}, -\ans{\eta+\gamma+1},\skipsym,\ldots,\skipsym) \nonumber\\
&- g(\ans{1},\ldots,\ans{\eta+\gamma}, \ans{\eta+\gamma+1}+1,\skipsym,\ldots,\skipsym) = 0.
\label{eq:lineareqn_subset_proof2}
\end{align}

The proof now proceeds via an induction on the quantity $({\numgold}-\eta-\gamma-1)$. We begin with the case of $({\numgold}-\eta-\gamma-1)={\numgold}-1$ which implies $\eta=\gamma=0$. In this case~\eqref{eq:lineareqn_subset_proof1} simplifies to
\begin{align*}
(1-\minbelief) g(\ans{1},\skipsym,\ldots,\skipsym) + \minbelief & g(-\ans{1},\skipsym,\ldots,\skipsym) = g(\ans{1}+1,\skipsym,\ldots,\skipsym).
\end{align*}
Applying the expansion of function $g$ in terms of function $\paysymbol$ from~\eqref{eq:skip_notation_subsample} for some $\ans{1} \in [\numchoices-1]$ gives
\begin{align*}
&(1-\minbelief) \left(c_1 \pay{\ans{1},\skipsym,\ldots,\skipsym}+c_2\pay{\skipsym,\skipsym,\ldots,\skipsym}\right) + \minbelief \left(c_1 \pay{-\ans{1},\skipsym,\ldots,\skipsym}+c_2 \pay{\skipsym,\skipsym,\ldots,\skipsym}\right) \nonumber\\
&\qquad\qquad\qquad\qquad\qquad\qquad\qquad\qquad = c_1 \pay{\ans{1}+1,\skipsym,\ldots,\skipsym}+c_2 \pay{\skipsym,\skipsym,\ldots,\skipsym}
\end{align*}
for constants $c_1>0$ and $c_2>0$ that respectively represent the probabilities that the first question is picked and not picked in the set of ${\numgold}$ gold standard questions. Cancelling out the common terms on both sides of the equation, we get the desired result
\begin{align*}
&(1-\minbelief) \pay{\ans{1},\skipsym,\ldots,\skipsym} + \minbelief \pay{-\ans{1},\skipsym,\ldots,\skipsym} = \pay{\ans{1}+1,\skipsym,\ldots,\skipsym}.
\end{align*}
Next, we consider the case when $({\numgold}-\eta-\gamma-1)$ questions are skipped in the gold standard, and assume that the result is true when more than $({\numgold}-\eta-\gamma-1)$ questions are skipped in the gold standard. In~\eqref{eq:lineareqn_subset_proof2}, the functions $g$ decompose into a sum of the constituent $\paysymbol$ functions. These constituent functions $\paysymbol$ are of two types: the first where all of the first $(\eta+\gamma+1)$ questions are included in the gold standard, and the second where one or more of the first $(\eta+\gamma+1)$ questions are not included in the gold standard. The second case corresponds to situations where there are more than $({\numgold}-\eta-\gamma-1)$ questions skipped in the gold standard and hence satisfies our induction hypothesis. The terms corresponding to these functions thus cancel out in the expansion of~\eqref{eq:lineareqn_subset_proof2}. The remainder comprises only evaluations of function $\paysymbol$ for arguments in which the first $(\eta+\gamma+1)$ questions are included in the gold standard. Since the last $(N-\eta-\gamma-1)$ questions are skipped by the worker, the remainder evaluates to
\begin{align}
&(1-\minbelief) c_3 \pay{\ans{1},\ldots,\ans{\eta+\gamma},\ans{i},\skipsym,\ldots,\skipsym} + \minbelief c_3 \pay{\ans{1},\ldots,\ans{\eta+\gamma},-\ans{i},\skipsym,\ldots,\skipsym} \nonumber\\
&\qquad\qquad\qquad\qquad\qquad\qquad\qquad\qquad\qquad\qquad\qquad= c_3 \pay{\ans{1},\ldots,\ans{\eta+\gamma},\ans{i}+1,\skipsym,\ldots,\skipsym}
\end{align}
for some constant $c_3>0$. Dividing throughout by $c_3$ gives the desired result.

Finally, the arguments above hold for any permutation of the first ${\numgold}$ questions, thus completing the proof.
\end{proof}

\subsection{Proof of Theorem~\ref{thm:possible_everyoption}: Mechanism under alternative formulation}
Without loss of generality, assume that $a=0$ and $b=1$ since in our setting, the property of incentive compatibility is invariant to any constant shift and positive scale of the payment.

First consider the case of $\numques = \numgold = 1$. Suppose without loss of generality that the worker's beliefs for the $\numchoices$ options are $p_1, \ldots , p_{\numchoices}$. It is easy to verify that the expected payment $\Epay_{\mbox{sup}}$ when the worker selects the options $\{o_1,\ldots,o_m\}$, for some $m$, equals 
\begin{align*}
\numchoices \thresholdEvery +  \sum_{i=1}^{\numchoices} (p_{o_i} - \thresholdEvery).
\end{align*}
It follows that the payment is strictly maximized when the worker selects all options whose beliefs are greater than $\numchoices$, given the assumption that none of the beliefs exactly equals $\thresholdEvery$.

The arguments above complete the proof for the case $\numques = \numgold = 1$. The extension to $\numques \geq \numgold \geq 1$ follow in a manner identical to the analogous extension in the proof of Theorem~\ref{thm:working}.

The proof of Corollary~\ref{cor:multiplicative_everyoption} follows in an identical fashion.

\subsection{Proof of Theorem~\ref{thm:impossible_everyoption}: Negative results under alternative formulation}
We present the results of uniqueness and impossibility respectively. We will let $\paysymbol$ denote any incentive compatible mechanism.

\subsubsection{Part A: Uniqueness}
Consider any $m \in \{1,\ldots,\selmax-1\}$. Consider the set of beliefs $\probopt{1}{} = \thresholdEvery + \delta$, $\probopt{2}{} = \cdots = \probopt{m+1}{} = \frac{1 - \thresholdEvery - \delta}{m}$ and $\probopt{m+2}{}=\cdots=\probopt{\numchoices}{}=0$, for some value of $\delta$ in the neighborhood of $0$. For the values of $m$ under consideration, one can verify that $\thresholdEvery < \frac{1 - \thresholdEvery}{m} < 1$. Consequently, there exists some value $\delta_{\max}>0$ such that for every $\delta \in [-\delta_{\max}, \delta_{\max}]$ we have $0 \leq \thresholdEvery + \delta \leq 1$ and $\thresholdEvery < \frac{1 - \thresholdEvery - \delta}{m} \leq 1$. In order to achieve the stated goal, we would thus require to incentivize the worker to select options $1$ through $(m+1)$ if $\delta>0$, and select options $2$ through $(m+1)$ if $\delta < 0$. The mechanism $\paysymbol$ therefore must satisfy the pair of inequalities
\begin{align*}
\pay{m+1} \overset{\delta < 0}{\underset{\delta > 0}{\lessgtr}} (1 - \thresholdEvery - \delta) \pay{m} + (\thresholdEvery + \delta) \pay{-m}.
\end{align*}
Since the right hand side of the expression above is linear in $\delta$ but the left hand side is a constant, we must have
\begin{align}
\pay{m+1} = (1 - \thresholdEvery) \pay{m} + \thresholdEvery \pay{-m} \qquad \mbox{for all $m \in \{1,\ldots,\selmax-1\}$}.
\label{eq:every_induction_1}
\end{align}
We will return to this equation later.

Next consider any $m \in \{1,\ldots,\selmax-2\}$. Consider the set of beliefs $\probopt{1}{} = \thresholdEvery + \delta$, $\probopt{2}{} = \thresholdEvery + \delta$, $\probopt{3}{} = \cdots = \probopt{m+2}{} = \frac{1 - 2\thresholdEvery - 2\delta}{m}$ and $\probopt{m+3}{}=\cdots=\probopt{\numchoices}{}=0$, for some value of $\delta$ in the neighborhood of $0$. For the values of $m$ under consideration, one can verify that $\thresholdEvery < \frac{1 - 2\thresholdEvery}{m} < 1$. Consequently, there exists some value $\delta_{\max}>0$ such that for every $\delta \in [-\delta_{\max}, \delta_{\max}]$ we have $0 \leq \thresholdEvery + \delta \leq 1$ and $\thresholdEvery < \frac{1 - 2\thresholdEvery - 2\delta}{m} \leq 1$. In order to achieve the stated goal, we would thus require to incentivize the worker to select options $1$ through $(m+2)$ if $\delta>0$, and select options $3$ through $(m+2)$ if $\delta < 0$. The mechanism $\paysymbol$ thus must satisfy
\begin{align*}
\pay{m+2} \overset{\delta < 0}{\underset{\delta > 0}{\lessgtr}} (1 - 2\thresholdEvery - 2\delta) \pay{m} + (2\thresholdEvery + 2\delta) \pay{-m}.
\end{align*}
Since the right hand side of the expression above is linear in $\delta$ but the left hand side is a constant, we must have
\begin{align}
\pay{m+2} = (1 - 2\thresholdEvery) \pay{m} + 2\thresholdEvery \pay{-m} \qquad \mbox{for all $m \in \{1,\ldots,\selmax-2\}$}.
\label{eq:every_induction_2}
\end{align}

It follows from~\eqref{eq:every_induction_1} and~\eqref{eq:every_induction_2} that the values of $\pay{m}$ for every $m \in \{-(\selmax - 1),\ldots,-1,1,\ldots,\selmax-2\}$ can be expressed in terms of a linear combination of $\pay{\selmax}$ and $\pay{\selmax-1}$. We will now prove that the same holds true for $\pay{-\selmax}$ and $\pay{0}$ as well, whenever these quantities are defined.

The quantity $\pay{-\selmax}$ is defined only when $\selmax < \numchoices$. The reason is that when $\selmax = \numchoices$, $\pay{-\selmax} = \pay{-\numchoices}$ corresponds to a scenario where all the options are selected and the correct option is not, which is impossible. Now consider the set of beliefs $\probopt{1}{} = \thresholdEvery + \delta$, $\probopt{2}{} = \cdots = \probopt{\selmax}{} = \frac{1 - \thresholdEvery - \delta - \epsilon}{\selmax - 1}$, $\probopt{\selmax + 1}{}= \epsilon$, and $\probopt{\selmax + 2}{}=\cdots=\probopt{\numchoices}{}=0$, for some values of $\epsilon \geq 0$ and $\delta$ in the neighborhood of $0$. From the definition of $\selmax$, one can easily verify that $\thresholdEvery < \frac{1 - \thresholdEvery - \epsilon}{\selmax - 1} < 1$ whenever $\selmax > 1$. Consequently, there exist some values $\delta_{\max}>0$ and $\epsilon_{\max} \in (0,\thresholdEvery)$ such that for every $\delta \in [-\delta_{\max}, \delta_{\max}]$ and for every $\epsilon \in [0, \epsilon_{\max}]$, we have $0 \leq \thresholdEvery + \delta \leq 1$ and when $\selmax > 1$, we also have $\thresholdEvery < \frac{1 - \thresholdEvery - \delta - \epsilon}{\selmax-1} \leq 1$. In order to achieve the stated goal, we would thus require to incentivize the worker to select options $1$ through $\selmax$ if $\delta>0$, and select options $2$ through $\selmax$ if $\delta < 0$. The mechanism $\paysymbol$ therefore must satisfy
\begin{align*}
(1-\epsilon) \pay{\selmax} + \epsilon \pay{-\selmax} \overset{\delta < 0}{\underset{\delta > 0}{\lessgtr}} (1 - \thresholdEvery - \delta - \epsilon) \pay{\selmax - 1} + (\thresholdEvery + \delta + \epsilon) \pay{-(\selmax-1)}.
\end{align*}
Since the right hand side of the expression above is linear in $\delta$ but the left hand side does not depend on$\delta$, we must have
\begin{align*}
(1-\epsilon) \pay{\selmax} + \epsilon \pay{-\selmax} = (1 - \thresholdEvery - \epsilon) \pay{\selmax - 1} + (\thresholdEvery + \epsilon) \pay{-(\selmax-1)}.
\end{align*}
Since this equation must be true for every $\epsilon \in [0, \epsilon_{\max}]$, we must have
\begin{align*}
-\pay{\selmax} + \pay{-\selmax} = - \pay{\selmax - 1} + \pay{-(\selmax-1)}.
\end{align*}
Thus the term $\pay{-\selmax}$, whenever applicable, can also be written as a linear combination of $\pay{\selmax}$ and $\pay{\selmax-1}$. 

The quantity $\pay{0}$ is defined only when $\thresholdEvery > \frac{1}{\numchoices}$. The reason is that when $\thresholdEvery \leq \frac{1}{\numchoices}$, it is mathematically impossible for the beliefs for all the $\numchoices$ options to be less than or equal to $\thresholdEvery$ (recall our assumption that no belief equals exactly $\thresholdEvery$). Now consider the set of beliefs $\probopt{1}{} = \thresholdEvery + \delta$, $\probopt{2}{} = \cdots = \probopt{\numchoices}{} = \frac{1 - \thresholdEvery - \delta}{\numchoices -1}$, for some value of $\delta$ in the neighborhood of $0$. One can verify that in this case of $\thresholdEvery > \frac{1}{\numchoices}$, it must be that $0 < \frac{1 - \thresholdEvery}{\numchoices -1} < \thresholdEvery$. Consequently, there exists some value $\delta_{\max}>0$ such that for every $\delta \in [-\delta_{\max}, \delta_{\max}]$, we have $0 \leq \thresholdEvery + \delta \leq 1$ and $0 \leq \frac{1 - \thresholdEvery - \delta}{\numchoices -1} < \thresholdEvery$. In order to achieve the stated goal, we would thus require to incentivize the worker to select option $1$ if $\delta>0$, and select no options if $\delta < 0$. The mechanism $\paysymbol$ therefore must satisfy
\begin{align*}
(\thresholdEvery + \delta) \pay{1} + (1 - \thresholdEvery  - \delta) \pay{-1} \overset{\delta < 0}{\underset{\delta > 0}{\lessgtr}} \pay{0}.
\end{align*}
Since the left hand side of the expression above is linear in $\delta$ but the right hand side is a constant, we must have
\begin{align*}
\thresholdEvery \pay{1} + (1 - \thresholdEvery) \pay{-1} = \pay{0}.
\end{align*}
Thus the term $\pay{0}$, whenever applicable, can also be written as a linear combination of $\pay{\selmax}$ and $\pay{\selmax-1}$. 

From the arguments above, we get that the design of $\paysymbol$ has only two degrees of freedom. Given that our claim is only up to some shift and scale, the claim is proved.

\subsubsection{Part B: Impossibility}
Let us first prove the result for the case of $\numques = \numgold = 1$. The result of part A of Theorem~\ref{thm:impossible_everyoption} implies that if there exists an incentive compatible mechanism for this setting, then the mechanism must be that of Mechanism~\ref{algo:incentive_every} up to a constant shift and positive scale. Consider a worker with the belief $\probopt{1}{} = 1 - \thresholdEvery$, $\probopt{2}{} = \thresholdEvery$ and $\probopt{3}{} = \cdots \probopt{\numchoices}{} = 0$. Since $\thresholdEvery < \frac{1}{2}$, under an incentive compatible mechanism, the expected payment must be strictly larger if the worker selects only option $1$ as compared to the expected payment when the worker selects options $1$ and $2$. However, one can compute that under Mechanism~\ref{algo:incentive_every}, the expected payment in the two cases is identical. It follows that under any possible incentive-compatible mechanism, the expected payment must be identical in the two following two actions of the worker (a) selecting only option $1$, and (b) selecting options $1$ and $2$. It follows that no mechanism is incentive compatible.

We now move on to the general case of $\numques \geq \numgold \geq 1$. Consider a worker who knows the answers to questions $2$ through $\numques$ with a belief of $1$ in each case. Suppose that for each of these $(\numques-1)$ questions, this worker selects the respective options that she thinks are correct. We are now left with the first question. Letting $X$ denote a random variable representing the evaluation of the worker's response to the first question, the expected payment from the worker's point of view is
\begin{align*}
\frac{\numgold}{\numques} \expect{ \pay{X, 1,\ldots,1} } + (1-\frac{\numgold}{\numques}) \pay{1,\ldots,1}.
\end{align*} 
The expectation in the first term is taken with respect to the randomness in $X$. Defining 
\begin{align*}
\tilde{\paysymbol}(X) \defn \frac{\numgold}{\numques} \pay{X, 1,\ldots,1} + (1-\frac{\numgold}{\numques}) \pay{1,\ldots,1},
\end{align*} 
and applying the same arguments to $\tilde{\paysymbol}$ as those for $\paysymbol$ for the case of $\numques=\numgold=1$ above gives the desired contradiction. This completes the proof.

\end{document}